\documentclass[letter,prd,aps,twocolumn,floatfix,superscriptaddress]{revtex4}

\usepackage{amssymb,amsmath,graphicx}
\usepackage{hyperref}
\usepackage[usenames,dvipsnames]{color}
\usepackage{array}
\makeatletter
\renewcommand*{\p@section}{\S\,}
\renewcommand*{\p@subsection}{\S\,}

\makeatother

\def\jnl@style{\it}
\def\aaref@jnl#1{{\jnl@style#1}}

\def\aaref@jnl#1{{\jnl@style#1}}

\def\aj{\aaref@jnl{AJ}}                   
\def\araa{\aaref@jnl{ARA\&A}}             
\def\apj{\aaref@jnl{ApJ}}                 
\def\apjl{\aaref@jnl{ApJ}}                
\def\apjs{\aaref@jnl{ApJS}}               
\def\ao{\aaref@jnl{Appl.~Opt.}}           
\def\apss{\aaref@jnl{Ap\&SS}}             
\def\aap{\aaref@jnl{A\&A}}                
\def\aapr{\aaref@jnl{A\&A~Rev.}}          
\def\aaps{\aaref@jnl{A\&AS}}              
\def\azh{\aaref@jnl{AZh}}                 
\def\baas{\aaref@jnl{BAAS}}               
\def\jrasc{\aaref@jnl{JRASC}}             
\def\memras{\aaref@jnl{MmRAS}}            
\def\mnras{\aaref@jnl{MNRAS}}             
\def\pra{\aaref@jnl{Phys.~Rev.~A}}        
\def\prb{\aaref@jnl{Phys.~Rev.~B}}        
\def\prc{\aaref@jnl{Phys.~Rev.~C}}        
\def\prd{\aaref@jnl{Phys.~Rev.~D}}        
\def\pre{\aaref@jnl{Phys.~Rev.~E}}        
\def\prl{\aaref@jnl{Phys.~Rev.~Lett.}}    
\def\pasp{\aaref@jnl{PASP}}               
\def\pasj{\aaref@jnl{PASJ}}               
\def\qjras{\aaref@jnl{QJRAS}}             
\def\skytel{\aaref@jnl{S\&T}}             
\def\solphys{\aaref@jnl{Sol.~Phys.}}      
\def\sovast{\aaref@jnl{Soviet~Ast.}}      
\def\ssr{\aaref@jnl{Space~Sci.~Rev.}}     
\def\zap{\aaref@jnl{ZAp}}                 
\def\nat{\aaref@jnl{Nature}}              
\def\iaucirc{\aaref@jnl{IAU~Circ.}}       
\def\aplett{\aaref@jnl{Astrophys.~Lett.}} 
\def\apspr{\aaref@jnl{Astrophys.~Space~Phys.~Res.}}
\def\bain{\aaref@jnl{Bull.~Astron.~Inst.~Netherlands}} 
\def\fcp{\aaref@jnl{Fund.~Cosmic~Phys.}}  
\def\gca{\aaref@jnl{Geochim.~Cosmochim.~Acta}}   
\def\grl{\aaref@jnl{Geophys.~Res.~Lett.}} 
\def\jcp{\aaref@jnl{J.~Chem.~Phys.}}      
\def\jgr{\aaref@jnl{J.~Geophys.~Res.}}    
\def\jqsrt{\aaref@jnl{J.~Quant.~Spec.~Radiat.~Transf.}}
\def\memsai{\aaref@jnl{Mem.~Soc.~Astron.~Italiana}}
\def\nphysa{\aaref@jnl{Nucl.~Phys.~A}}   
\def\physrep{\aaref@jnl{Phys.~Rep.}}   
\def\physscr{\aaref@jnl{Phys.~Scr}}   
\def\planss{\aaref@jnl{Planet.~Space~Sci.}}   
\def\procspie{\aaref@jnl{Proc.~SPIE}}   

\begin{document}

\date{\today}
\title{Effects of the microphysical Equation of State in
the mergers of magnetized Neutron Stars With Neutrino Cooling}

\author{Carlos Palenzuela}
\affiliation{Departament de F\'isica, Universitat de les Illes Balears and Institut d'Estudis Espacials de Catalunya, Palma de Mallorca, Baleares
E-07122, Spain}
\author{Steven L. Liebling}
\affiliation{Department of Physics, Long Island University, Brookville, New York 11548, USA}
\author{David Neilsen}
\affiliation{Department of Physics and Astronomy, Brigham Young University, Provo, Utah 84602, USA}
\author{Luis Lehner}
\affiliation{Perimeter Institute for Theoretical Physics,Waterloo, Ontario N2L 2Y5, Canada}
\author{O. L. Caballero}
\affiliation{Department of Physics, University of Guelph, Guelph, Ontario N1G 2W1, Canada}
\author{Evan O'Connor}
\affiliation{Department of Physics, North Carolina State University, Raleigh, North Carolina 27695, USA; Hubble Fellow}
\author{Matthew Anderson}
\affiliation{Pervasive Technology Institute, Indiana University, Bloomington, IN 47405, USA}


\begin{abstract}
We study the merger of binary neutron stars using different realistic, microphysical
nuclear equations of state, as well as incorporating magnetic field
and neutrino cooling effects. In particular, we concentrate on the influence of the 
equation of state on the gravitational wave signature and also on its role, in combination
with cooling and electromagnetic effects, in determining the properties of the
hypermassive neutron star resulting from the merger, the production of neutrinos, and
the characteristics of ejecta from the system. 
The ejecta we find are consistent with other recent studies 
that find soft equations of state produce more ejecta than stiffer equations of state. Moreover,
the degree of neutron richness increases for softer equations of state. 
In light of reported kilonova observations (associated to GRB~130603B and GRB~060614) and
the discovery of relatively low abundances of heavy, radioactive
elements in deep sea deposits (with respect to possible production via supernovae),
we speculate that a soft EoS might be preferred---because of
 its significant production of sufficiently neutron rich ejecta---if such events are driven by binary neutron star mergers.
We also find that realistic magnetic field strengths, obtained with a sub-grid model tuned to
capture magnetic amplification via the Kelvin-Helmholtz instability at merger,
are generally too weak to affect the gravitational wave signature post-merger within
a time scale of $\approx 10$~ms 
but can have subtle 
effects on the post-merger dynamics.
\end{abstract}

\maketitle

\tableofcontents

\section{Introduction}\label{introduction}

Binary neutron star systems~(BNS) represent perhaps the most exciting and anticipated
source for upcoming gravitational wave~(GW) observations.
In contrast to other expected binary sources such as neutron star--black hole binaries,  BNS
have been observed and these observations lead to reasonably robust lower bounds on the
expected GW detections~\cite{2010CQGra..27q3001A}. In addition, neutron stars, being
made of matter and supporting strong magnetic fields, are more
complicated than black holes, especially during merger.  The coalescence of neutron
stars, being a highly dynamical violent event, disrupts the stars, powers electromagnetic events,
and produces large neutrino bursts.
Other effects may produce precursor signals prior to merger
through either the interaction of stellar
magnetospheres~\cite{Palenzuela:2013hu,Palenzuela:2013kra,Ponce:2014sza} or 
crust cracking~\cite{Tsang:2011ad}.

The merger produces a differentially rotating, 
hot massive neutron star (MNS) with a strong neutrino luminosity.
Such mergers are suspected engines of short, gamma ray bursts~(sGRB),
arguably the most spectacular electromagnetic counterpart
to a strong GW event.
Binaries composed of a black hole and neutron star may also power sGRB, but,
in order for such binaries to do so, either low
mass ratios or black holes with significant spin are required,
e.g.~\cite{Shibata:2006bs,Etienne:2008re,Chawla:2010sw,Foucart:2013psa}.
In both systems, the merger may produce 
a significant neutron rich ejecta which can trigger kilonova events~\cite{Li:1998bw} 
(see also~\cite{Kulkarni:2005jw,Metzger:2008av,Kasen:2014toa}). 
Remarkably, a recent observation of a sGRB led to a follow-up
observation in the infrared indicative of a kilonova event~\cite{2013Natur.500..547T,Berger:2013wna}.
This observation provides further support for the connection between non-vacuum binary mergers
with sGRBs~\cite{Berger:2013jza}.

Moving beyond the broad expectations of BNS mergers sketched above, the detailed dynamics of such mergers depend
on a number of parameters, such as
the stellar  masses, equation of state, and magnetization, to name but a few.
Fortunately, the combined efforts by the astrophysics and numerical relativity communities are gradually 
refining our understanding of these systems and the observational opportunities they present. For instance, simulations
have: 
indicated that
the merger will give rise to a black hole promptly when the total mass of the system 
$M_{\rm total} \gtrsim 2.8 M_{\odot}$ (e.g.~\cite{Sekiguchi:2012uc,Fryer:2015uia});
provided an approximate range for the final spin of the black hole for cases that do collapse~\cite{2013PhRvD..88b1501K};
indicated that the collision can produce magnetic field strengths in the resulting MNS
that range from  magnetar levels to higher~\cite{Price:2006fi,B_fields_are_us,2013ApJ...769L..29Z};
and
contributed gravitational wave templates to aid in data analysis efforts (e.g.~\cite{Read:2009yp,Baiotti:2011am,Damour:2012yf}).

Most recently, a number of research avenues are attracting considerable interest.
A significant effort is examining possible electromagnetic counterparts, of which
the aforementioned sGRB are just one example
(see e.g.~\cite{Lehner:2011aa,Kyutoku:2012fv,2012ApJ...746...48M,Palenzuela:2013hu,Andersson:2013mrx,2014arXiv1410.8560R}.
Another is the study of binaries with more realistic equations of state~\cite{Read:2009yp,Neilsen:2014hha,Read:2013zra,2015arXiv150206660S}. 
In particular, the hope is that
information about the true equation of state for NS interiors can be obtained
from observations.  Related to such efforts is an intense interest 
in ejecta and outflows~\cite{2013PhRvD..87b4001H,2015arXiv150206660S}, because of their
connection to possible
kilonova events and their role in explaining the origin of r-process 
elements~\cite{2015NatCo...6E5956W}. 
Neutron star binary characteristics, including ejecta, have also been actively studied 
within Newtonian models~\cite{Piran:2014wpa}, where the dynamics sometimes exhibit 
subtle, but important, differences with respect to analogous systems in full general relativity.
Additionally, the role of a magnetic field in BNS mergers is also under study. Potentially,
an organized post-merger field may play a crucial role in the generation of
a jet were a binary to power a sGRB. Since it is generally expected that
neutron stars have some magnetization, determining what electromagnetic effects
could arise is important.

Here we explore the dynamics of binary neutron star systems with
increased realism: 
we study  stars described by different realistic equations of state; we account 
for neutrino cooling effects through a leakage algorithm; and we consider the role of
magnetization in the post-merger dynamics.
Details of our numerical implementation can be found in previous
work~\cite{Anderson:2006ay,Anderson:2008zp,Liebling:2010bn,Palenzuela:2013hu,Neilsen:2014hha}. 
We summarize details in section~II, present results in section~III, and discuss these results in section~IV.
We defer to appendices a description of the primitive solver and convergence tests.

\section{Numerical Implementation}
\label{sec:implementation}
Below we briefly summarize: 
(i)~the evolution equations for the spacetime and the
magnetized fluid; 
(ii)~the prescriptions employed for the microphysical equations of state and the neutrino 
cooling scheme; 
and
(iii)~a description of the quantities employed to analyze the dynamics.
Full details of our implementation 
are described in~\cite{Neilsen:2014hha}. We adopt geometrized units 
where $G=c=M_\odot = 1$, although some results are reported more naturally 
in physical cgs units.

\subsection{Evolution equations}

The Einstein equations in the presence of both matter and radiation can be
written as
\begin{equation}
G_{ab} = 8 \pi (T_{ab} + {\cal R}_{ab} ),
\label{Eins1}
\end{equation}
where $G_{ab}$ is the Einstein tensor, ${\cal R}_{ab}$ is the contribution from 
the radiation field, and $T_{ab}$ is the stress energy tensor of a magnetized, perfect fluid.

We solve the Einstein equations by adopting a 3+1 decomposition in terms
of a spacelike foliation. The hypersurfaces that constitute this
foliation are labeled by a time coordinate $t$ with unit normal $n^a$ and
endowed with spatial coordinates $x^i$. We write the spacetime metric as
\begin{equation}
ds^2 = -\alpha^2\,dt^2 
+ \gamma_{ij}\left(dx^i + \beta^i\,dt\right)\left(dx^j + \beta^j\,dt\right),
\end{equation}
with $\alpha$ the lapse function, $\beta^i$ the shift vector and
$\gamma_{ij}$ the metric on spatial hypersurfaces.
We write the Einstein equations in terms of the BSSN-NOK 
formalism~\cite{1987PThPS..90....1N,1995PhRvD..52.5428S,1999PhRvD..59b4007B,
lousto}, supplemented with appropriate gauge conditions; the ``1+log'' slicing
and the $\Gamma$-driver shift conditions, as detailed in~\cite{Neilsen:2014hha}.
In the problem of interest, radiation energies and stresses are much smaller
than matter energies, or $|{\cal R}^{ab}| \ll |T^{ab}|$. 
This observation, together with the fact that
a neutrino leakage scheme can not possibly treat the radiation
stress tensor completely consistently, allows us to ignore 
${\cal R}_{ab}$ as a source term for the Einstein equations. 

We model the matter as a magnetized perfect fluid with a stress energy 
tensor $T_{ab}$ given by
\begin{equation}
T_{ab} = h u_a u_b + P g_{ab}  
+ F_{ac} F^c_b - \frac{1}{4} g_{ab} F_{cd} F^{cd}\, ,
\end{equation}
where $h$ is the fluid's {\it total} enthalpy $h \equiv  \rho (1+\epsilon) + P$,
and $\{\rho,\epsilon,Y_e,u^a,P\}$ are the rest mass energy density, 
specific internal energy, electron fraction (describing the relative abundance of electrons compared to the total number of baryons), four-velocity, and pressure of the fluid, 
respectively. Magnetic effects are included in the Faraday tensor,
 $F_{ab}$, where a factor $1/\sqrt{4 \pi}$ has been absorbed in its definition.
Given an equation of state $P=P(\rho,\epsilon,Y_e)$ and a relativistic Ohm's law, the equations determining the magnetized
matter dynamics are obtained from conservation laws. 
We adopt
the ideal magnetohydrodynamics~(MHD) approximation, 
which states that the fluid is described by
an isotropic Ohm's law with perfect conductivity so that the electric
field vanishes in the fluid's frame $F_{ab} u^b =0$, because
it provides a simple, yet realistic, model that keeps the number of
fields needed to describe the electromagnetic effects to a minimum
(in contrast to, say, a resistive MHD approach~\cite{Palenzuela:2008sf,2013MNRAS.431.1853P}).

The equations of motion consist of the following conservation laws
\begin{align}
\nabla_a T^a_b &= {\cal G}_b\, , \label{eq:DT}\\
\nabla^a (T_{ab} n^b) &= 0\, , \label{eq:DTN} \\
\nabla_a (Y_e \rho u^a) &= \rho R_Y\, , \label{eq:divYe}\\
\nabla_a {}^* F^{ab} &= 0 \label{eq:DF}\,.
\end{align}
The sources ${\cal
  G}_a$~($\equiv -\nabla_c {\cal R}^c_{a}$) and $R_Y$ are the radiation
four-force density and lepton sources which are determined within the
leakage scheme. 
These equations
are conservation laws for the stress-energy tensor, matter, lepton number, and Maxwell tensor, respectively. 
 Notice that, in the absence of lepton source terms ($R_Y=0$),
Eq.~(\ref{eq:divYe}) becomes a conservation law for leptons, similar to 
the baryon conservation law, i.e., $Y_e$ is a mass scalar. 
These conservation equations combined with the Einstein equations 
Eq.~(\ref{Eins1}) comprise the complete set of equations we employ to describe
the systems of interest.

The matter equations of motion in Eqs.~(\ref{eq:DT})--(\ref{eq:DF}) are written in balance law form
\begin{equation}
\partial_t {\bf u} + \partial_i{\bf f}^i\left({\bf u}\right) = {\bf s}({\bf u})
\end{equation}
by defining a set of {\it conservative} variables ${\bf u}$. The primitive
(or physical) quantities are recovered from the conservative ones by solving 
a system of non-linear equations, described in  detail in Appendix~\ref{appendix:primsolver}.
Due to the inability of current fluid codes to evolve very rarefied matter,
we impose a floor on the density, so that a tenuous atmosphere fills the domain outside the stars with 
constant density of $6 \times 10^{5}$~g/cm$^3$.

We use finite difference techniques on a regular, 
Cartesian grid to discretize
the system~\cite{SBP2,SBP3}. 
The geometric fields are discretized
using a fourth order accurate scheme satisfying the summation by
parts rule, while a High-Resolution Shock-Capturing method based on the HLLE
flux formulae with PPM reconstruction are used to discretize the fluid and
the electromagnetic variables. The fluid equations are written in finite
difference form using a third-order CENO method.
The time evolution of the resulting equations is performed by using a third order accurate Runge-Kutta scheme~\cite{Anderson:2006ay,Anderson:2007kz}. 
To ensure sufficient resolution is available in an efficient manner, we employ
adaptive mesh refinement~(AMR) via the HAD computational infrastructure that
provides distributed, Berger-Oliger style AMR~\cite{had_webpage,Liebling} with
full sub-cycling in time, together with an improved treatment of artificial
boundaries~\cite{Lehner:2005vc}.

\subsection{Equations of State and neutrino transport}

We use finite-temperature equations of state based on several
relativistic mean field models of the nuclear interaction to model
both neutron star matter and the hot, lower density material present
during and after the merger. These different nuclear interactions give
different cold neutron star structures and correspondingly different
neutron radii and maximum masses. The use of a microphysical EoS over
a simple polytrope is also needed to model neutrino interactions as
the neutrino emission, absorption, and scattering rates depend
sensitively on the matter density, temperature and composition.  We use
publicly available EoS tables from www.stellarcollapse.org described
in~\cite{O'Connor:2009vw}.  We have rewritten some of the library
routines for searching the table to make them faster and more robust.

The leakage scheme seeks to account for (1) changes to the
(electron) lepton number and (2) the loss of energy from the emission
of neutrinos.  Because the dynamical timescale for the
post-merger is relatively short,
radiation momentum transport and diffusion are expected to be
subleading effects. As is standard, we consider three species of neutrinos:
$\nu_e$ for electron neutrinos, $\bar{\nu_e}$ for electron antineutrinos,
and $\nu_x$ for both tau and muon neutrinos and their respective antineutrinos.
Our scheme is based on the open-source neutrino leakage
scheme from Ref.~\cite{O'Connor:2009vw}, also available at
www.stellarcollapse.org. In particular, the leakage scheme provides the
fluid rest frame energy sink ${\cal Q}$ and lepton sink/source $R_Y$ due to
neutrino processes.  $R_Y$ is the source term for a scalar quantity
and therefore is the same in all frames.  We express the source term
for the energy and momentum in an arbitrary frame as
\begin{equation}
{\cal G}_{a} = {\cal Q} u_a \, .
\end{equation}
Since the effect of neutrino pressure is small~\cite{Deaton:2013sla} and difficult
to accurately capture with a neutrino leakage scheme, we ignore its
contribution in the fluid rest frame. 
As mentioned, the details of our implementation can be found
in Ref.~\cite{Neilsen:2014hha}, which also describes a novel and efficient
calculation of the optical depth.

\subsection{Analysis}
We compute several quantities to analyze the quality of the solution as well as the 
dynamics and main characteristics of the system.
A useful quantity to assess the correctness of the numerical solution is the total baryonic mass 
in the domain
\begin{equation}
  M_b = \int \rho W \sqrt{\gamma} \, dx^3,
\label{eq:baryonic_mass}
\end{equation}
where $W$ is the Lorentz factor. Because this mass should remain constant, we monitor it and also
study its convergence in Appendix~\ref{appendix:convergence}.

Among the three possible signals produced by a BNS,
the cleanest signal is provided by gravitational waves,
which are potentially observable by the upcoming aLIGO/VIRGO to a distance of
$\approx 450$~Mpc. 
We extract this signal from our simulations 
by computing the Newman-Penrose scalar $\Psi_4$, which is directly related to the metric perturbation 
components by $ \Psi_4  = \ddot{h}_{+} - i \ddot{h}_{x}$ in the appropriate Bondi frame~\cite{Newman:1961qr,Newman:1981fn} (c.f.~\cite{lehnermoreschi}). We compute $\Psi_4$ on several spheres centered on the origin 
and with radii of $\{ 225,\, 300,\, 375\} \rm{km}$, i.e, far from the sources, in the wave zone.
Performing the calculation on multiple spheres allows various consistency checks to confirm the validity of the signal.
As is customary, we expand $\Psi_4$ in terms of spin-weighted spherical harmonics (with spin weight $s=-2$)
\begin{equation}
 r \Psi_4 (t,r,\theta,\phi) = \sum_{l,m} \Psi_4^{l,m} (t,r) \, {}^{-2}Y_{l,m} (\theta,\phi).
\label{eq:psi4}
\end{equation}
We also extract the instantaneous GW angular frequency from the
primary  ($l=m=2$) mode through the relation
\begin{equation}
 \omega = -\frac{1}{2} \rm{Im} \left[ \frac{\dot{\Psi}_4^{2,2}}{\Psi_4^{2,2}} \right].
\label{eq:omega}
\end{equation}

The high cost of BNS simulations makes computing the full
waveform relevant for matched-filtering analysis for data obtained by detectors such as aLIGO/VIRGO
in full relativity impractical. However, for 
larger separations (low frequencies), 
post-Newtonian (PN)
analyses provide excellent approximations.  Therefore, our  simulations concentrate primarily on the merger stage, and hybrid
waveforms are constructed by matching the simulation waveforms to 
PN waveforms. Among the available PN waveforms, the Taylor-T4 approximation
has proven quite convenient for binary black hole systems~\cite{2007PhRvD..76l4038B}, and has recently been expanded
to account for tidal effects~\cite{Vines:2011ud} (cf.~\cite{Hotokezaka:2013mm}). Our simulations span the last
4--6 orbits (depending on the EoS) and can be matched to the PN waveforms (by suitably blending
over a cycle) to construct a complete waveform within aLIGO's frequency window. This allows us to calculate the gravitational
wave power as a function of  frequency and contrast it with the expected noise curve of aLIGO.\\

Another possible signal is provided by neutrinos produced by the
shock-heated material during merger and afterward. 
Provided the event takes place relatively nearby, such neutrinos would be observable
 in current detectors such as Super-Kamiokande~\cite{SuperK} and IceCube~\cite{IceCube}
as well as future detectors such as Hyper-Kamiokande~\cite{HyperK}. We
extract the neutrino luminosity and compute from it the distance at which a detection could take place.\\

Electromagnetic signals are also possible and would provide complementary information about
the system (and with the possibility of covering a much wider set of frequencies due to the mature stage of detectors
in the electromagnetic band). To extract such information, reasonable models connecting possible observations with
particular characteristics of the systems are required.
The details of such models generally rely on different assumptions and,
in many cases, involve longer timescales than those that can be reasonably probed within a full general relativity simulation.
However, one can explore the level to which consistent conditions are provided for such schemes.
In particular, the lifetime of the MNS, 
the timescale for possible black hole formation, and the characteristics of any material outflow 
are key physical parameters (see e.g. 
and references cited therein ~\cite{Kelley:2012tc,Metzger:2011bv,Andersson:2013mrx,Lehner:2014asa}). 
Thus, we monitor the behavior of the formed MNS---e.g. central density and
multipolar structure---as well as the properties of matter within our simulations, supplemented with reasonable approximations
to connect with possible phenomena at longer timescales.

We assess the dynamics of all stellar material, at some representative
times after the merger, by assuming it moves along geodesics. We compute its specific energy (as measured at infinity)
($\tilde E \equiv -u_t - 1$)  and consider the material unbound if $\tilde E > 0$. We then produce histograms displaying
various quantities which characterize the state of the bound and unbound material for the
different EoS.
In particular, to  generate a histogram characterizing some quantity $x$ (e.g. temperature), we compute
a series of histogram heights, $h_i$, for bins extending from $b_i \rightarrow b_{i+1}$ (e.g. $10$ MeV to $20$ MeV) as an integral over the 
whole space
$
h_i = \int \, \Theta(x-{b_i}) \Theta(b_{i+1}-x) \rho dV,
$
where $\Theta(x)$ is the Heaviside function.
In practice a single sweep over the domain suffices because, at each gridpoint, the integrand gets added to just one bin.

\section{Numerical results}
\label{sec:numericalresults}

\begin{figure}[h]
\centering
\includegraphics[width=8.5cm,angle=0]{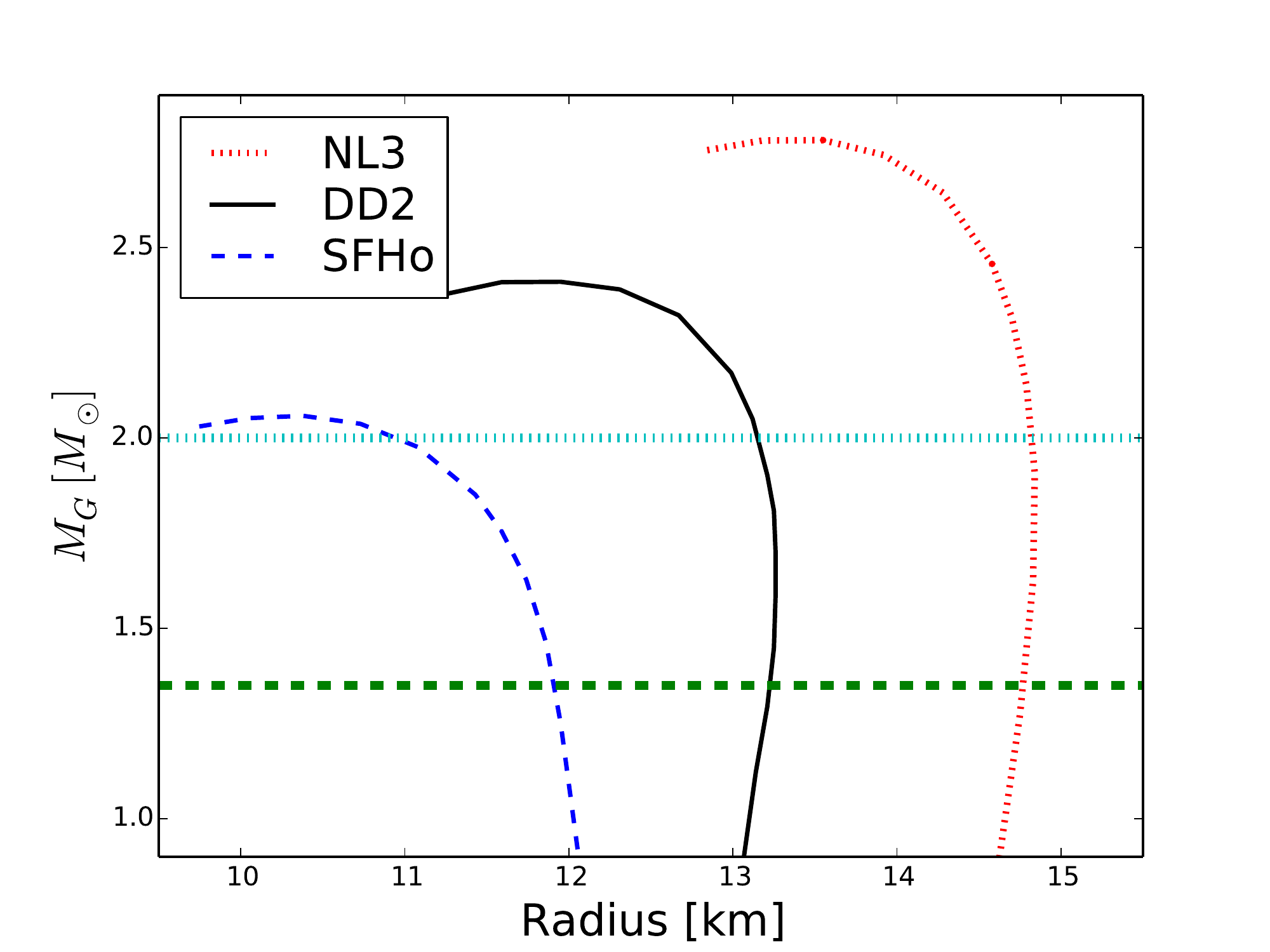}
\caption{Gravitational mass as a function of the circumferential radius for the three
different microphysical EoS considered here. The horizontal, light blue line 
 shows the maximum mass  astrophysically measured
for a neutron star~\cite{Demorest:2010bx,Antoniadis:2013pzd} and the horizontal, green line indicates the mass considered in this paper.
These curves were generated with the
{\sc Magstar} solver, part of the {\sc Lorene}
package~\cite{lorene}.
} 
\label{fig:mass_radius}
\end{figure}

We concentrate on equal-mass binaries with individual gravitational masses given by $1.35 M_{\odot}$, a value 
consistent with current observations~\cite{doi:10.1146/annurev-nucl-102711-095018}. 
We consider three distinct microphysical EoS: SFHo~\cite{2013ApJ...774...17S}, DD2~\cite{2012ApJ...748...70H}
and NL3~\cite{2012ApJ...748...70H}. 
These three EoS produce cold neutron stars whose radii cover a large
range of values. 
In particular, the SFHo EoS gives the smallest radii ($\approx 12$ km for a mass of $1.35 M_{\odot}$) among the
three and is therefore considered a {\em soft} EoS. 
The NL3 EoS yields a large radius ($\approx 15$ km for the same mass) and is considered {\em stiff}.
The DD2 EoS leads to an intermediate radius ($\approx 13$ km).
We display the gravitational mass of isolated neutron stars for each EoS 
in Fig.~\ref{fig:mass_radius}.
A horizontal, green line at $1.35 M_{\odot}$ shows the stars considered here and its intersection
with the three curves indicates the three different radii. All three EoS can produce neutron stars with a mass
of at least 2\,$M_\odot$, consistent with neutron star mass
observations \cite{Demorest:2010bx,Antoniadis:2013pzd}.

We concentrate on the late stages of the coalescence, roughly the last $4$--$6$ orbits (depending on EoS), of
equal mass, irrotational binary neutron stars. 
The physical parameters of the binaries and our grid setup are summarized in Table~\ref{table:equal_mass}.
To monitor convergence of the solutions, we evolved the binary described by the DD2 EoS with three different resolutions obtaining convergent results (see Appendix~\ref{appendix:convergence}). 
For the two other EoS, we evolved only two different resolutions confirming that the differences
between the two were consistent with those of the DD2 case.
(For reference, unless otherwise specified, we set $t=0$ as the time when all binaries are separated by $45$km.)

As mentioned, we include 
the effects of neutrino cooling and magnetic field in the dynamics. To reduce the significant computational 
overhead that these introduce,
neutrino cooling is not calculated until a few milliseconds before merger when matter heats up
and neutrino effects become significant. 
Likewise, we give the stars a poloidal magnetic field  (with maximum strength $10^{13}$G) 
a few milliseconds prior to merger. 
This strategy allows us to
consider both the role of neutrino production and
stellar magnetization in an efficient manner 
because neither cooling (the stars are cold prior to merger)
nor magnetic field effects~\cite{Ioka:2000yb,Anderson:2008zp} affect the dynamics prior to merger.

\begin{table*}[t]\centering
\begin{tabular}{|l|l|l|l|l|l|l|l|l|l|}
\hline
EoS & $M_{0}^{\rm ADM} [M_{\odot}]$ & $J_{0}^{\rm ADM} [G M^2_{\odot}/c]$ & $m_{b}^{(i)} [M_{\odot}]$  & $r^{(i)}$ [km] & $\Omega_0$ [rad/s] &
$f_0^{\rm GW}$ [Hz]  & $M_{\rm eject} [M_{\odot}]$ & $\Delta x_{\rm min}$ [m] \\ \hline
 NL3  & 2.7  & 7.40 & 1.470  & 12.63 & 1778 & 566 & $1.6\times 10^{-6}$  & 230   \\ \hline
 DD2  & 2.7  & 7.39 & 1.485  & 10.88 & 1776 & 565 & $4.3\times 10^{-4}$  & 230 (275 \& 192)  \\ \hline
 SFHo & 2.7  & 7.38 & 1.498  & 9.47  & 1775 & 565 & $3.2\times 10^{-3}$  & 230 \\ \hline
\end{tabular}
\caption{Summary of the parameters of the binary systems considered in this work. The initial data were computed using the {\sc Bin star} solver from the {\sc Lorene} package~\cite{lorene}, by imposing beta-equilibrium and a constant temperature $T=0.02$ MeV. All the binaries start from an initial separation of $45$ km and the outer boundary is located at $750$ km: the ADM mass $M_{0}^{\rm ADM}$ and angular momentum $J_{0}^{\rm ADM}$ of the system, the baryon mass of each star $m_b^{(i)}$ and its coordinate radius $r^{(i)}$, the initial orbital angular frequency $\Omega_0$ of the system, the initial GW frequency $f_0^{\rm GW} \equiv \Omega_0/\pi$ and the best resolution $\Delta x_{\rm min}$ covering both the stars. For the DD2 EoS, higher and lower resolutions were carried out for
the study of convergence (see Appendix~\ref{appendix:convergence}).}
\label{table:equal_mass}
\end{table*}

\subsection{Gravitational Waves}

At large separations, all binaries follow the same trajectory since
the internal structure of the binary constituents only contributes 
at 5th PN order (see e.g.~\cite{2014grav.book.....P}). 
However, as the binary tightens,  
differences from a point-particle approximation arise and increase as the merger ensues.
Such differences are intrinsically related to the tidal deformability of the stars
which  is, in turn, directly tied to the EoS.

\begin{figure}[h]
\centering
\includegraphics[width=8.5cm,angle=0]{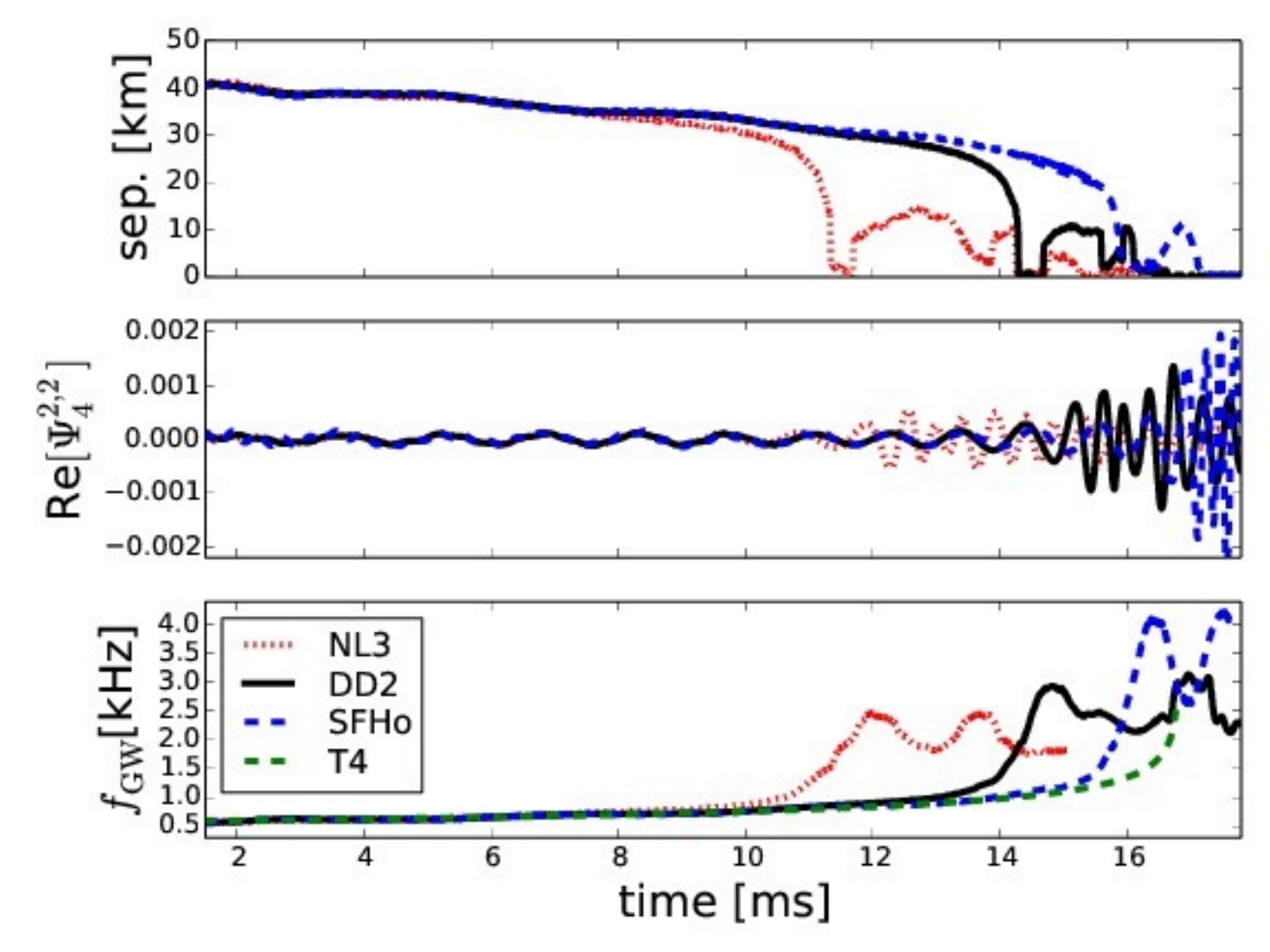}
\caption{ Orbital parameters displayed as a function of time for the three different EoS considered here. 
(Top) Separation between the stars, computed as the coordinate distance between the maximums of the density.
(Middle) Primary  gravitational wave mode $l=m=2$. 
(Bottom) The frequency of the primary GW mode where $f_{\rm GW} = \omega/\left(2\pi \right)$ (see Eq.~(\ref{eq:omega})). 
Also shown is the frequency obtained from the
Taylor T4 approximation. Note that differences among the EoS only appear at times near merger.
} 
\label{fig:separation_omega}
\end{figure}

We summarize the pre-merger dynamics of our three binaries in Fig.~\ref{fig:separation_omega}.
The (coordinate) separation, the GW signal $\Psi_4$ ($l=2,m=2$ component), and 
the instantaneous GW frequency, all as functions of time, are shown.
Also included in the bottom frame is the gravitational wave frequency as predicted
by the T4 PN approximation~\cite{2007PhRvD..76l4038B}. During the first eight milliseconds, all the binaries
agree with each other and with the T4 approximation. Only at times closer to merger do the binaries
begin to differentiate from each other and from the point-particle approximation.

\begin{figure}[h]
\centering
\includegraphics[width=8.5cm,angle=0,clip]{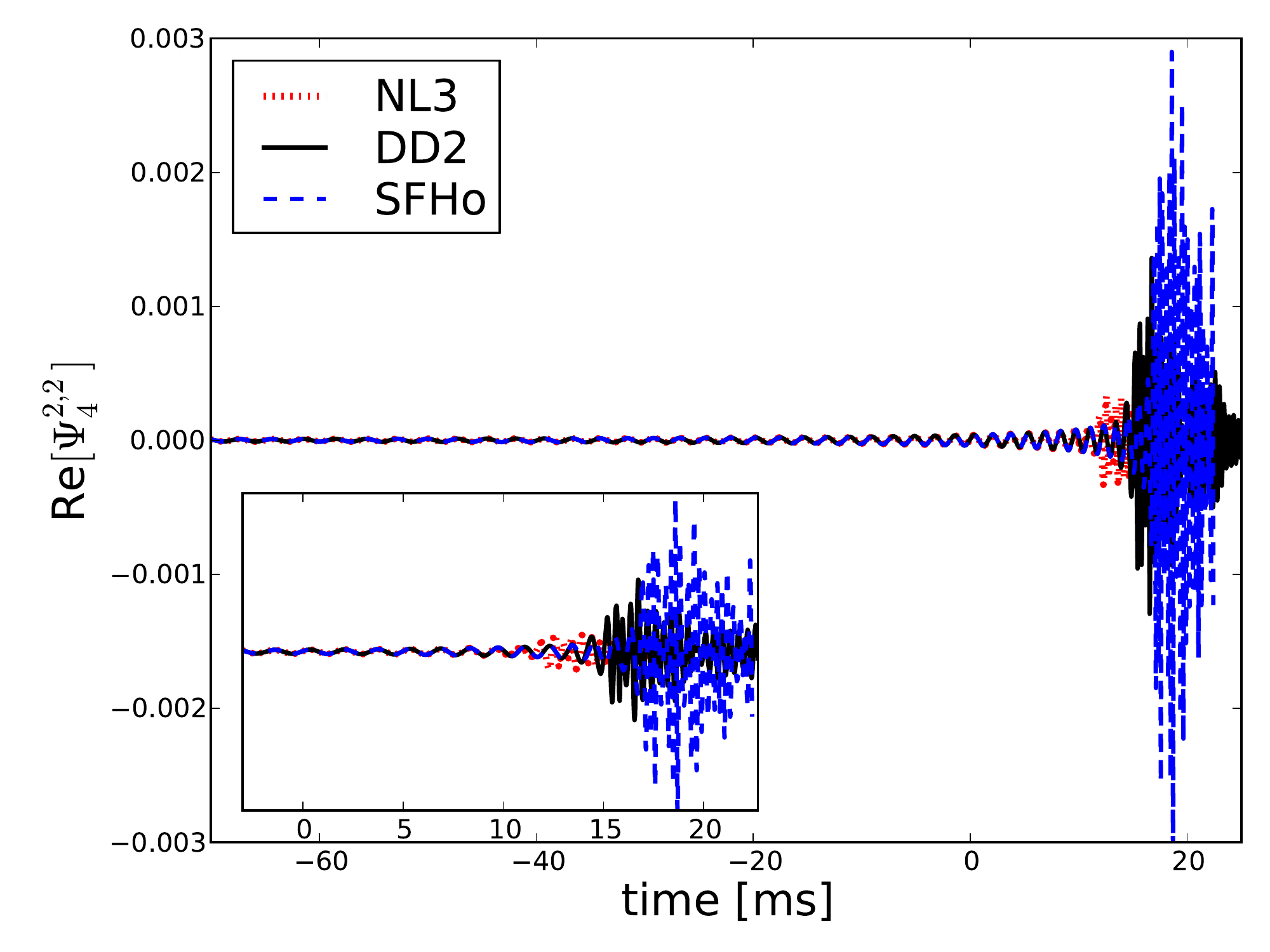}
\caption{Hybrid waveforms obtained by matching the PN waveform with the respective EoS waveform  from the
evolution of the binary. These waveforms are used for comparing signals with aLIGO noise curves
in Fig.~\ref{fig:strainligo}.
} 
\label{fig:wavetrain}
\end{figure}

The first EoS to show differences 
is NL3, followed by DD2, and then SFHo. This is the expected behavior as the NL3 EoS corresponds
to the largest neutron star radii, 
DD2 the intermediate and SFHo the smallest radii.  Tidal effects are proportional to $k_2 R^5$
(where $k_2$ is the tidal Love number and $R$ is the stellar radius), and therefore, for fixed mass, differences in radii are dominant.
We take advantage of this early agreement to compute a complete wavetrain by creating a hybrid of the PN and our simulated waveforms
which we show in Fig.~\ref{fig:wavetrain} and discuss later.
Of course, the merger itself is highly nonlinear, and 
so an examination of the merger waveforms does not involve any further comparison with PN results.

\begin{figure}[h]
\centering
\includegraphics[width=8.0cm,angle=0]{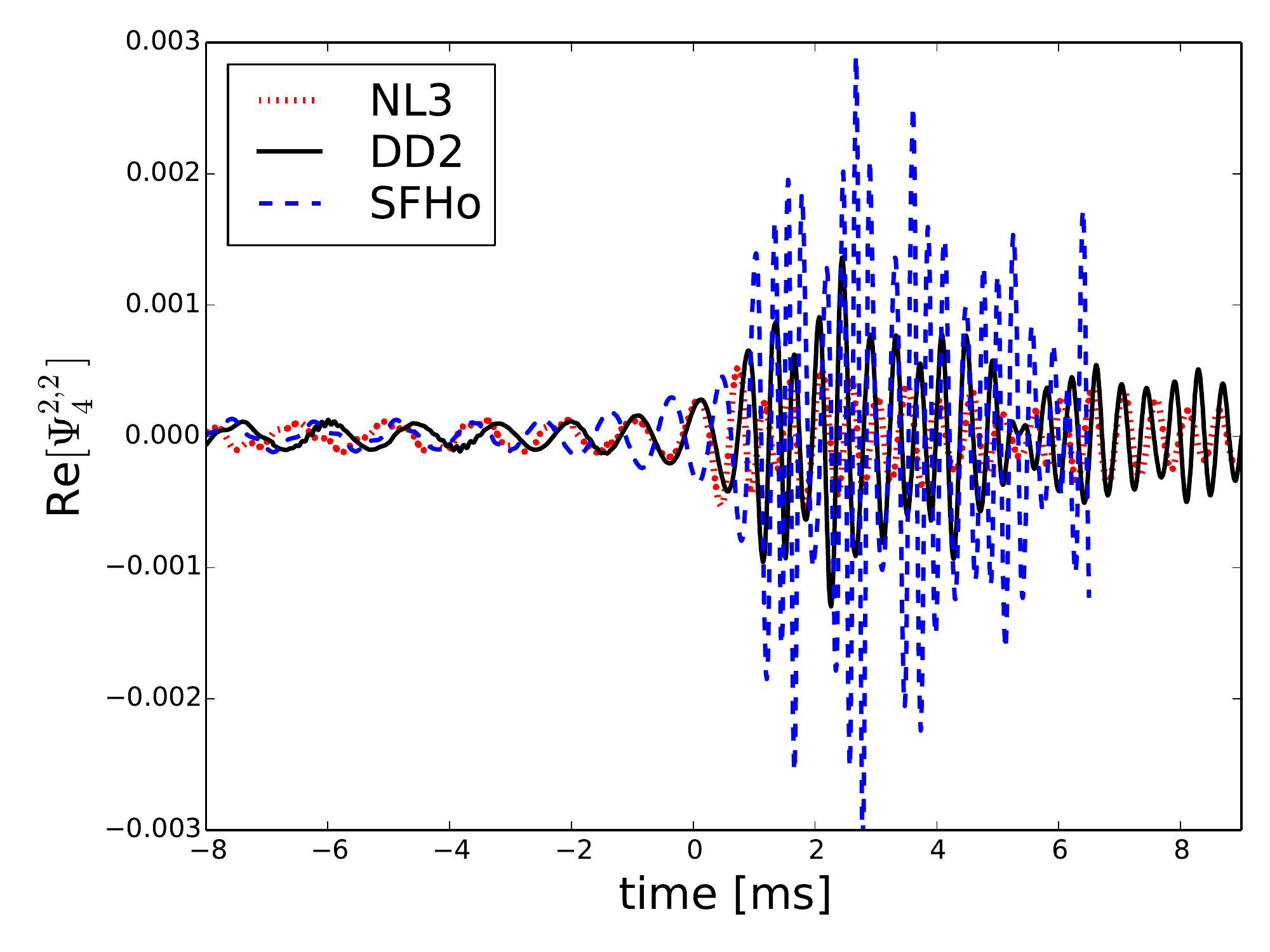}
\caption{ Comparison of the near-merger behavior of the GW signal, as indicated by the primary mode of
$\Psi_4$. The different signals have been shifted in time such that
$t=0$ corresponds to the time when the separation between the density maximums vanishes.
The significant differences among the signals  near merger indicate that
information about the internal stellar structure (e.g. details of the EoS, radius, compactness) is encoded within
the gravitational wave signature.
}
\label{fig:waves1}
\end{figure}

Considering now the merger, we shift times in each case such that
$t=0$ refers to the moment when the maximum density first occurs at the origin of the binary.
Fig.~\ref{fig:waves1} presents the gravitational waveforms with time measured from the
merger. 
Note that all three signals prior to merger (negative times) oscillate within a common amplitude envelope. This
behavior can be understood 
with an effective, and approximate, one-body problem point of view. 
In such a problem, the central object has a mass approximately given by the total mass of the system and radiative properties
can be captured by the infall of a particle of reduced mass $\mu= m_1 m_2/(m_1+m_2)$. Shifting time so that 
all cases merge at $t=0$ allows for identifying a natural common time, and phase differences
are a consequence of the particle being released at different times with respect to the shifted time.
Such an approach
has been at the core of a number of  effective prescriptions for capturing various aspects of the two body problem in general relativity (e.g.~\cite{Buonanno:1998gg,Buonanno:2007sv,Foucart:2012nc,Bernuzzi:2014owa}).

Fig.~\ref{fig:waves1} also
allows for a clear comparison of the strength of the gravitational waves and the waveform periods.
The SFHo EoS achieves the largest amplitude during merger as
a natural consequence of yielding  the smallest stellar radii. A smaller radius means that
the merger takes place deeper in the gravitational potential of the binary, and hence more
energy is released, resulting
in a more dynamical and violent collision. 
In contrast, the weakest peak-signal with longest period is  produced by the NL3 EoS with its large radii while the DD2 binary is
intermediate between the two extremes considered.

\begin{figure}[h]
\centering
\includegraphics[width=9.0cm,angle=0]{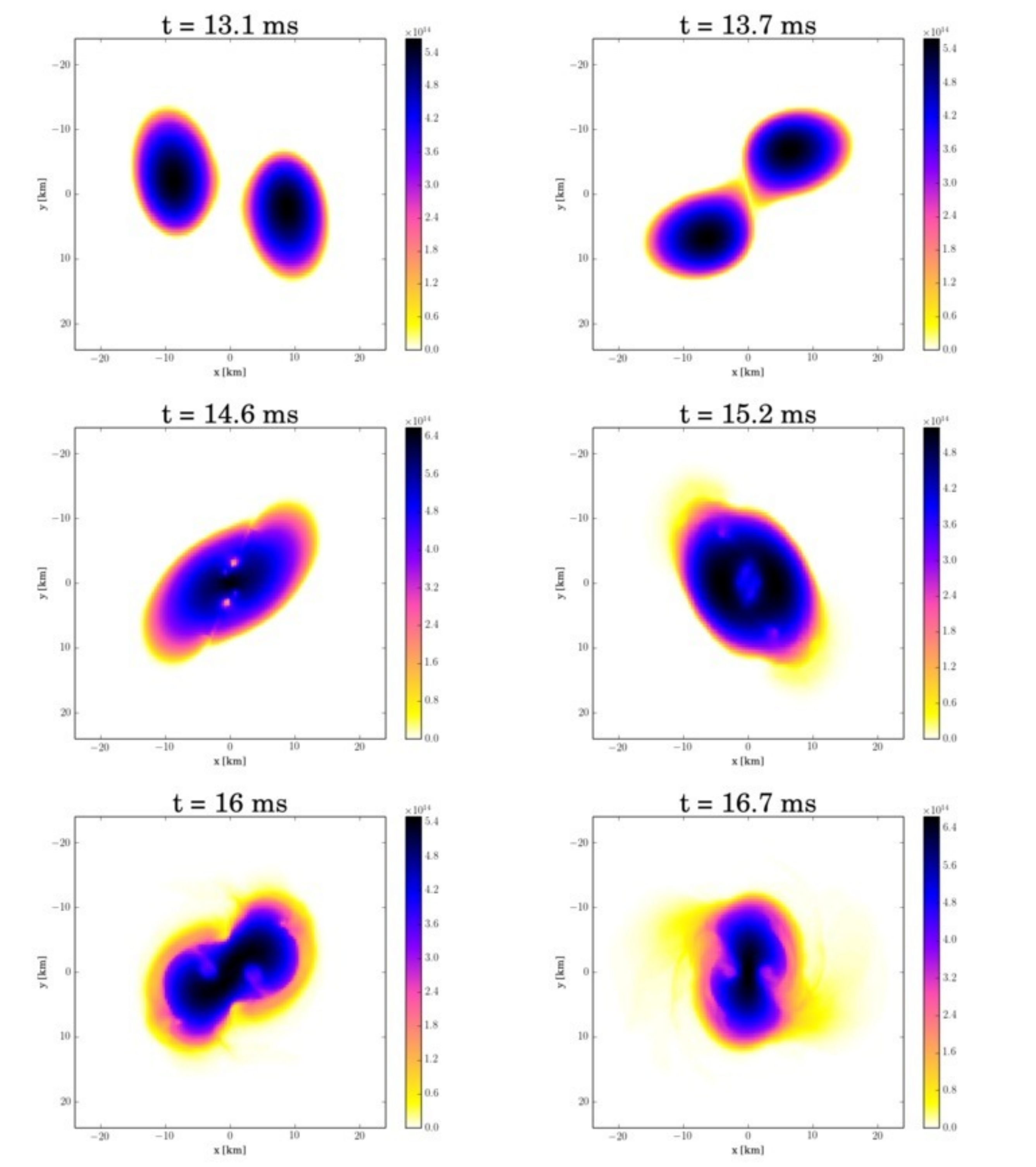}
\caption{ Density on the equatorial plane for the DD2 EoS. The stars merge at $t \approx 14.3$ ms. The first
two frames show the system pre-merger as tidal deformations arise. Also visible
in the final four frames is the rebound of the stellar material post-merger.
} 
\label{fig:density_2d}
\end{figure}

Qualitatively, the mergers for all three EoS proceed similarly.
The stellar orbit tightens as GW are emitted, and, as the
stars get closer, tidal deformations grow. Evidence for such deformations can be appreciated, for the DD2 case, in
the first two frames of Fig.~\ref{fig:density_2d}. (We caution however that coordinate effects are not taken into
account in the figure; observable effects of tidal deformations are better represented in the gravitational
waveforms obtained.) After the merger, strong radial oscillations in
the remnant lead to peak densities bouncing outward (with respect to the co-rotating frame), 
as shown in the last four frames of  Fig.~\ref{fig:density_2d}.
These bounces are also apparent in the separation plot in Fig.~\ref{fig:separation_omega}.
The oscillations
however are stronger---larger in amplitude and longer lasting---as the EoS becomes stiffer, 
with mergers at earlier times and smaller angular velocities 
(as can be appreciated in Figs.~\ref{fig:separation_omega} and~\ref{fig:waves1}). 

The resulting hot MNS will cool down as well as lose energy and angular momentum via emission of neutrinos 
and gravitational waves. Furthermore, its velocity distribution will be redistributed
through angular momentum transport by hydrodynamical and electromagnetic effects. Collapse to a
black hole will eventually take place if the mass of the MNS is higher than the maximum allowed mass for a given EoS. 
Inspection of Fig.~\ref{fig:mass_radius}, together with the observation that the 
amount of unbound mass is relatively small in all cases, indicates that
at least the SFHo case is expected to collapse to a black hole 
at some point (see also~\cite{2015arXiv150206660S,Fryer:2015uia}). Such an expectation arises naturally because the maximum mass allowed with this EoS for a non-rotating star
is  $\approx 2.05 M_{\odot}$, and, while rotation helps increase the maximum allowable mass, it can not do so beyond $\approx 20\%$~\cite{1994ApJ...424..823C}. Indeed, for the times studied in this work, the SFHo binary
collapses to a black hole $\approx 7$~ms after merger.

\begin{figure}[h]
\centering
\includegraphics[width=8.5cm,angle=0]{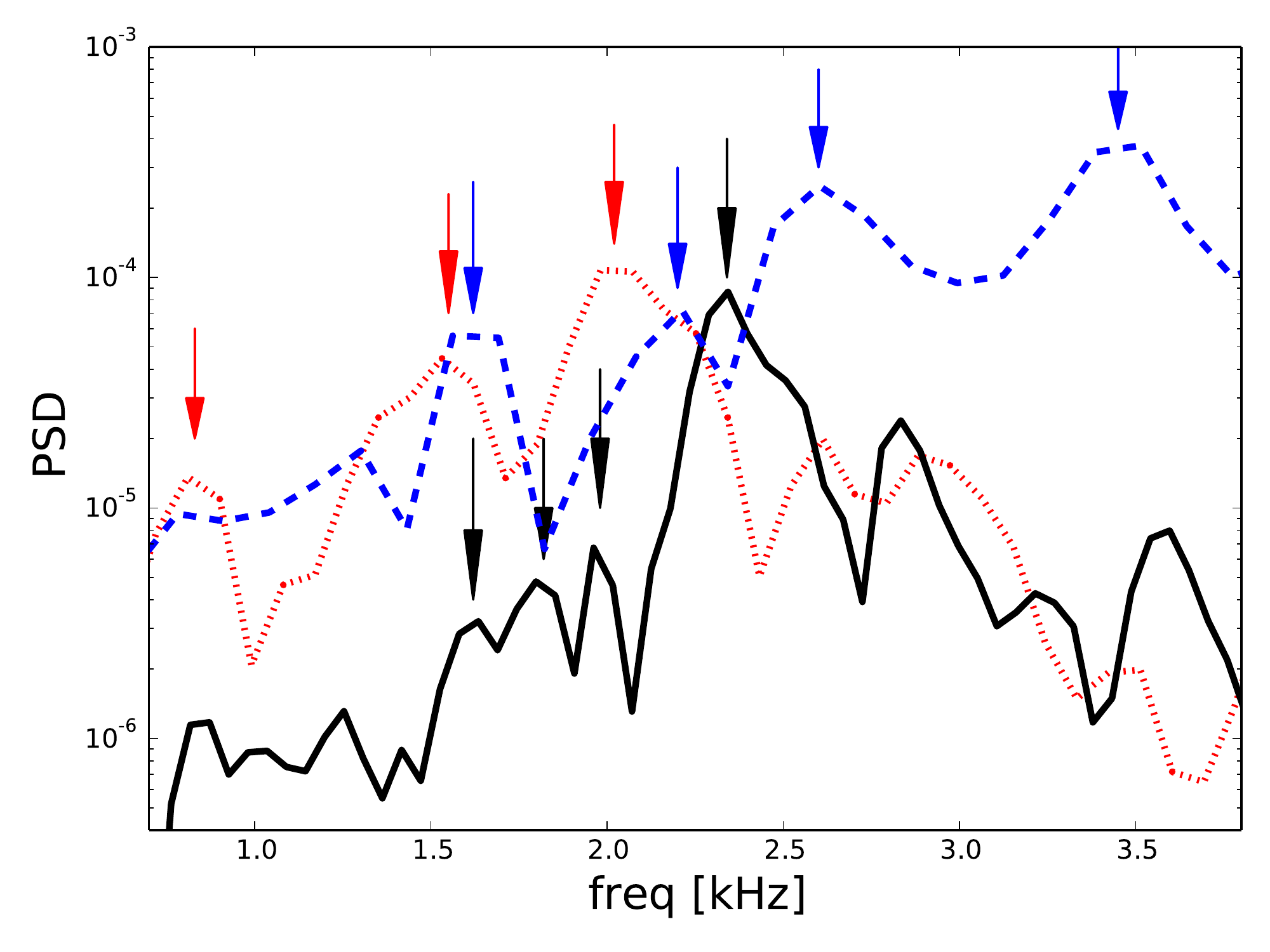}
\caption{ Power spectral densities~(PSD) of the gravitational waveforms after
the merger. The Fourier transforms display the main oscillation modes of the MNS.
The arrows indicate particular peaks with their corresponding frequencies
$f_i$ presented in Table~\ref{table:modes_HMNS}.
} 
\label{fig:eoswaveform_postmerger}
\end{figure}

Additional information about the properties of the neutron star are provided by the GW after the merger.
Fig.~\ref{fig:eoswaveform_postmerger} displays the power
spectral density of the Fourier transform for the main GW mode at this stage. The spectrum shows
several characteristic peaks. The dominant peak (the one with the most power), $f_{\rm peak}$,  
is associated with the rotational behavior and quadrupolar structure of the MNS.
As mentioned, a softer EoS results in a merger at higher frequency as the smaller radius stars
coalesce. This effect is apparent when comparing the spectra for the different EoS in
Fig.~\ref{fig:eoswaveform_postmerger}.
The peak frequencies are also presented in Table ~\ref{table:modes_HMNS}.

Beyond this qualitative discussion, the peak frequency is slightly higher than twice the orbital 
frequency at the time of merger. 
Since there will be further compression of the merged MNS, its rotational frequency 
will be higher than the frequency estimated by Kepler's law at the moment of contact. 
Nevertheless, the Keplerian orbital frequency can be used to estimate
$f_{\rm peak}$ as\footnote{An empirical fit to the peak
frequency can also be obtained as in~\cite{2015arXiv150203176B}}
\begin{equation}
f_{\rm peak}^i/f_{\rm peak}^j \approx (R_j/R_i)^{3/2},
\end{equation}
where $i$ and $j$ denote different binaries with stellar radii given by $R_i$ and $R_j$, respectively.
Interestingly,
several other subdominant peaks exist at lower frequencies that arise from oscillations of 
the MNS. 
All these peaks offer interesting future detectability prospects for 3rd generation 
detectors (aLIGO/VIRGO/KAGRA) 
that will have specifically tuned configurations to increase sensitivity in the higher
frequency range or even signal-stacking possibilities (e.g.~\cite{Abbott:2009zd}).

Recently, Ref.~\cite{2015arXiv150203176B} indicated that these low frequency, secondary modes
can be traced to two different mechanisms, although this is still a matter of debate (see for instance~\cite{2015PhRvD..91f4027K,2015PhRvD..91f4001T}). Using the same naming conventions of~\cite{2015arXiv150203176B}, the first, secondary peak
$f_{\rm 2-0}$ corresponds to a non-linear combination of the dominant quadrupolar mode with the quasi-radial oscillations of the star.
The other sub-dominant peak, $f_{\rm spiral}$, is produced by a strong spiral deformation induced at the merger time that lasts for
just a few rotational periods.
We calculate these frequencies for our simulations using the estimates in
Ref.~\cite{2015arXiv150203176B}, which are based on the compactness of the neutron stars, and include 
these in Table ~\ref{table:modes_HMNS}.
Comparing the frequencies extracted from our simulations with the estimates, we find
that the simulation values for both the dominant,
$f_1$, and secondary, $f_2$, peaks are consistent with $f_{\rm peak}$ and $f_{\rm spiral}$, respectively.
Also, the simulation frequencies $\{f_3,f_4\}$ are in reasonable agreement with $f_{2-0}$ 
for the softest and intermediate EoS (SFHo and DD2).

\begin{table}[t]\centering
\begin{tabular}{|l|l|l|l|l||l|l|l|l|}
\hline
EoS &  $f_{1}$ &  $f_{2}$  & $f_{3}$ & $f_{4}$ & $M/R$ 
&  $f_{\rm peak}$  & $f_{\rm spiral}$ & $f_{2-0}$ \\ \hline
\hline
NL3  & 2.0 & 1.5  & 0.8  & --   & 0.135 & 2.2  & 1.6  & 1.2  \\ \hline
DD2  & 2.3 & 2.0  & 1.8  & 1.62 & 0.15  & 2.6  & 1.9  & 1.5  \\ \hline
SFHo & 3.4 & 2.6  &  2.2 & 1.6  & 0.167 & 3.2 & 2.4  & 2.1 \\ \hline
\end{tabular}
\caption{Post-merger oscillation frequencies in kHz. The frequencies of the various peaks of the post-merger GW spectrum (see Fig.~\ref{fig:eoswaveform_postmerger}) are shown as 
$f_1, f_2, f_3,$ and $f_4$. The compactness of the neutron stars for each EoS (obtained from the initial data for an isolated star) along with the predicted peak frequencies from Ref~\cite{2015arXiv150203176B}.
The decent agreement between $f_1$ and $f_{\rm peak}$, between $f_2$ and $f_{\rm spiral}$, and between
either $f_3$ or $f_4$ with $f_{2-0}$
suggests consistency with the model presented in~\cite{2015arXiv150203176B}.
}
\label{table:modes_HMNS}
\end{table}

Finally, we return to the complete gravitational wavetrain. As discussed, the early
behavior is well approximated by the PN (Taylor T4) expansion which 
can be augmented to include leading order tidal effects~\cite{Vines:2011ud,Hotokezaka:2013mm}.
We thus obtain a more complete waveform by matching this augmented PN signal at early times to
our numerical signal before merger to form a hybrid waveform.
The matching is accomplished by choosing a cycle after initial numerical transients but still
well before merger. Within this cycle, a weighted average of the two waveforms is calculated
using 
a tanh function that transitions from $0$ to $1$. We show (a portion of) the 
hybrid waveforms resulting
from the three EoS in Fig.~\ref{fig:wavetrain}.

\begin{figure}[h]
\centering
\includegraphics[width=8.5cm,angle=0,clip]{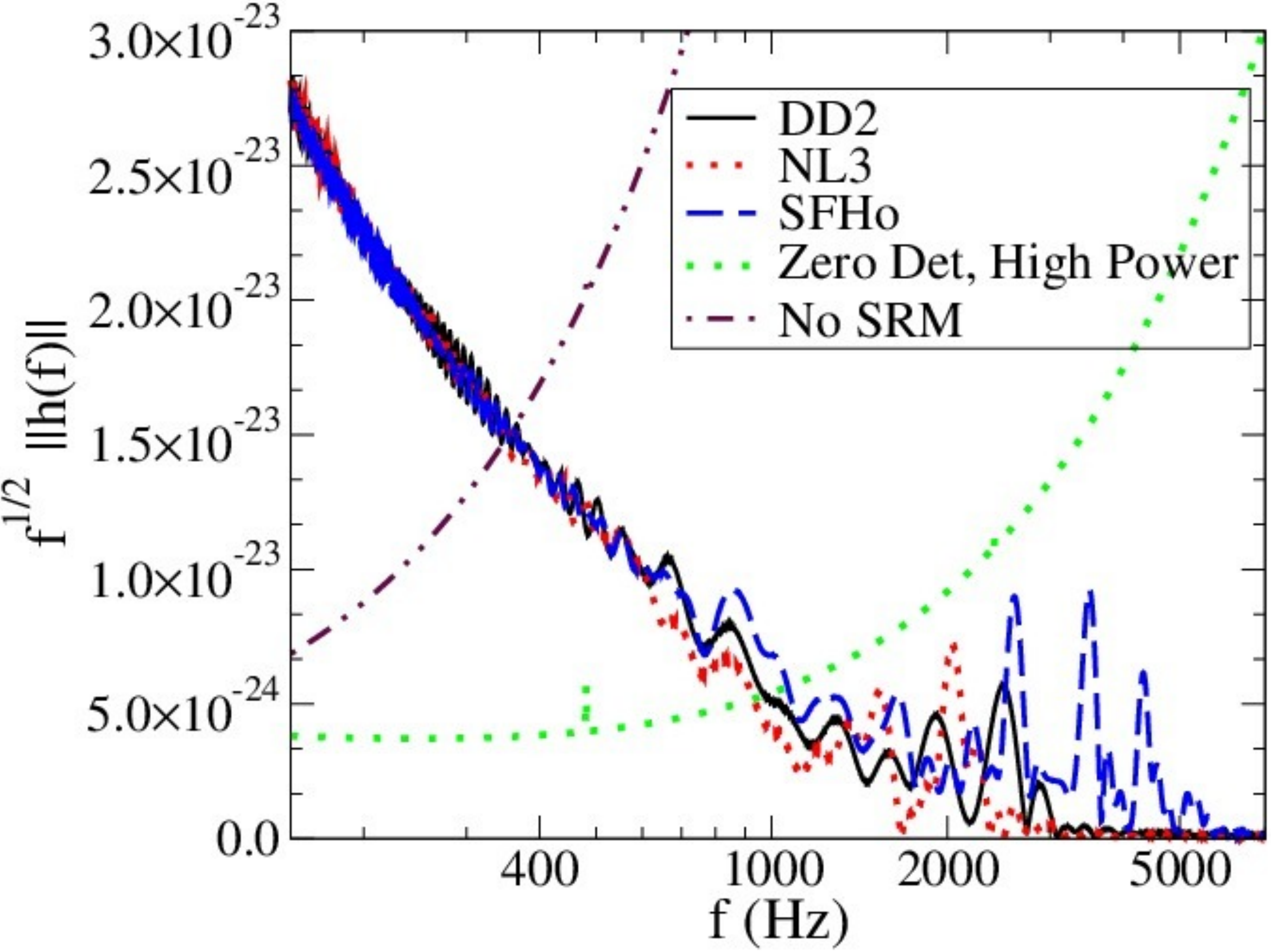}
\caption{Fourier  spectrum  of  the GW signals
for the three binaries. The dotted and dashed-dotted monotonic 
curves at high frequencies show
two fits to the noise power spectrum $\sqrt{S_h}$ of aLIGO (specifically the ``zero-detuned high-power'' and the
``No Signal Recycling Mirror''~(NSRM) cases, see~\cite{LIGOCURVE}).
} 
\label{fig:strainligo}
\end{figure}

Additionally, we compare the strength of the signals to the expected noise
curve of Advanced LIGO as illustrated in Fig.~\ref{fig:strainligo}, which 
presents the Fourier spectrum of the GW
(multiplied by $\sqrt{f}$) for binaries at $100$~Mpc versus frequency together
with two fits to the aLIGO Noise curve (the so called ``No Signal
Recycling mirror'' and the ``zero-detuned high-power case''). 
As mentioned, differences among
the waveforms arise only at high frequencies when tidal effects become relevant. The NL3 waveform
departs first from the other two cases because the larger stellar radii lead to stronger tidal effects.
The differences among the three EoS
arise below the aLIGO noise-curve in the case without a signal recycling mirror but could
be captured in the ``broadband'' (zero-detuned, high power case) configuration.

Finally, we can compute the ``distinguishability'' between the different waveforms 
(as described in e.g.~\cite{Read:2013zra}, see also~\cite{Owen:1995tm,Baumgarte:2006en,Read:2009yp})
using the ``zero-detuned high-power case'' noise curve in aLIGO. Without allowing for frequency shifts
(and not normalizing our waveforms), we find the values for 
$\delta h = \sqrt{\langle h_i-h_j,\, h_i-h_j\rangle}$ are given by
$\approx \{1.33,\, 1.36,\, 1.45\}$ when considering the pairs SFHo-DD2, SFHo-NL3 and DD2-NL3 respectively within
the frequency window $f \in[300,8000]$~Hz. Since all cases satisfy $\delta h>1$ the gravitational waves will
be distinguishable from each other by aLIGO with the ``zero-detuned high-power case'' configuration. 
In contrast, these values reduce to $\delta h \approx \{0.27,0.28,0.32\}$ for the ``No Signal Recycling mirror'' configuration.

\subsection{Matter dynamics: Outflow and Ejecta}
\label{sec:outflows}

\begin{figure}[h]
\centering
\includegraphics[width=9.2cm,angle=0,clip]{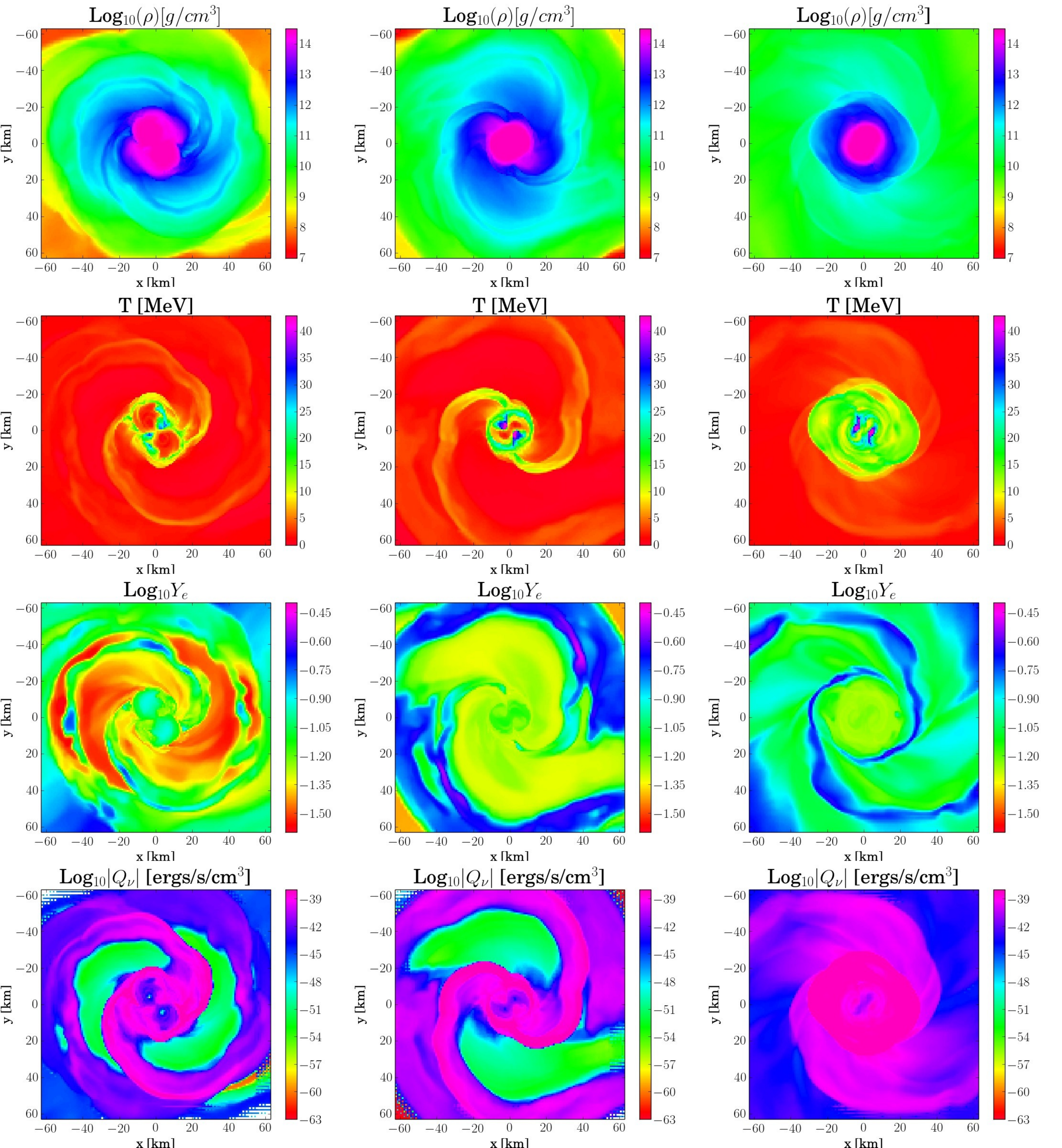}
\caption{Density, temperature, electron fraction and neutrino luminosity rate for the different EoS in the $z=0$ plane at roughly $t=3$~ms after the merger.
Shown are the NL3 EoS (left), DD2 EoS (middle), and SFHo (right).
} 
\label{fig:eos_2dplots_zplane}
\end{figure}

\begin{figure}[h]
\centering
\includegraphics[width=9.2cm,angle=0,clip]{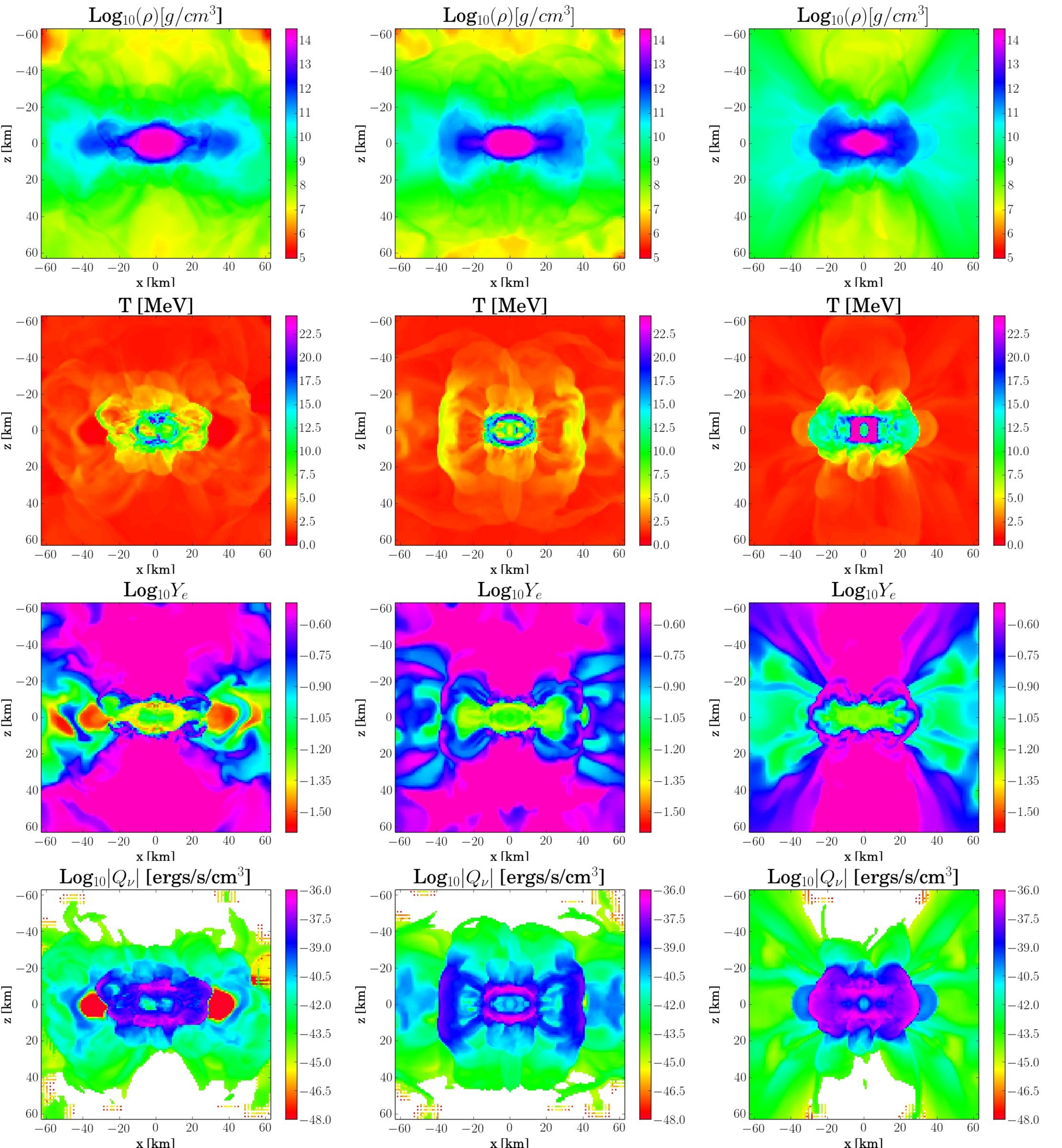}
\caption{Density, temperature, electron fraction, and neutrino luminosity rate for the different EoS in the $y=0$ plane at roughly $t=3$~ms after the merger.
Shown are the NL3 EoS (left), DD2 EoS (middle), and SFHo (right).
} 
\label{fig:eos_2dplots_xplane}
\end{figure}

Having presented characteristics of the GW produced mainly by the bulk motion of the stars, we now turn to
an analysis of the material outflow resulting from the collision. While such motion leads only to subleading
effects on the bulk dynamics (and hence on GW emission), the details of any outflow can be very important
for producing electromagnetic and neutrino signals.

In particular, among the outflow, material bound to the system either remains part of the
resulting MNS or otherwise contributes to an accretion disk which may potentially power a sGRB (see e.g.~\cite{Nakar:2007yr}).
On long time scales, such a disk can induce winds that produce optical electromagnetic
emission via the radioactive decay of r-process elements~\cite{Kasen:2014toa}.
Furthermore, the fallback of bound material may trigger the eventual collapse of the remnant
star to a black hole which, in turn, could power 
an intense burst of electromagnetic radiation~\cite{Lehner:2011aa,Murguia-Berthier:2014pta}.

Another particularly interesting mechanism powered by outflows is the proposed kilonova in which 
material ejected by the collision undergoes r-process nucleosynthesis. 
Being ejected from the core region, the behavior of this material and its consequences can be analyzed without worrying about the
rather involved behavior of the central region. Indeed, as discussed in~\cite{Li:1998bw}, a promising source of electromagnetic
signals is the radioactive decay of r-process elements formed in the ejecta
(see also, e.g.~\cite{Metzger:2008av}). Such signals have  specific characteristics observable
on the order of days after the merger~\cite{2013Natur.500..547T,Berger:2013wna}. Importantly, such characteristics
can have a rather tight dependence on the parameters of the binary and the EoS describing the neutron stars involved.

Given the importance of the outflow, in the following we compare the properties of both bound and unbound material among the three EoS.\\

We begin our analysis of the merger and postmerger by presenting snapshots of important quantities just a few milliseconds after merger. Figs.~\ref{fig:eos_2dplots_zplane} and~\ref{fig:eos_2dplots_xplane}
display the density, temperature, electron-fraction, and neutrino luminosity along the $z=0$ and $y=0$ planes for the three EoS.
The density snapshots show clear differences among the EOS, with the SFHo having a much more distributed, high density region
than the NL3 remnant. This difference can be understood in terms of the stiffness of the EoS. 
In particular, the SFHo, being the softest
of these EoS with the most compact stars, results in a more violent merger; the compact stars have further to travel to merge
and have higher impact velocities. The violence serves to distribute the stellar material more than is seen with the stiffer
EoS and, in particular, yields more ejecta. Likewise, the temperature of the material is generally higher for the softer EoS and, as a consequence, the neutrino
production is also higher (we discuss further neutrino-related aspects in the next subsection).

\begin{figure}[h]
\centering
\includegraphics[width=4.cm,angle=0]{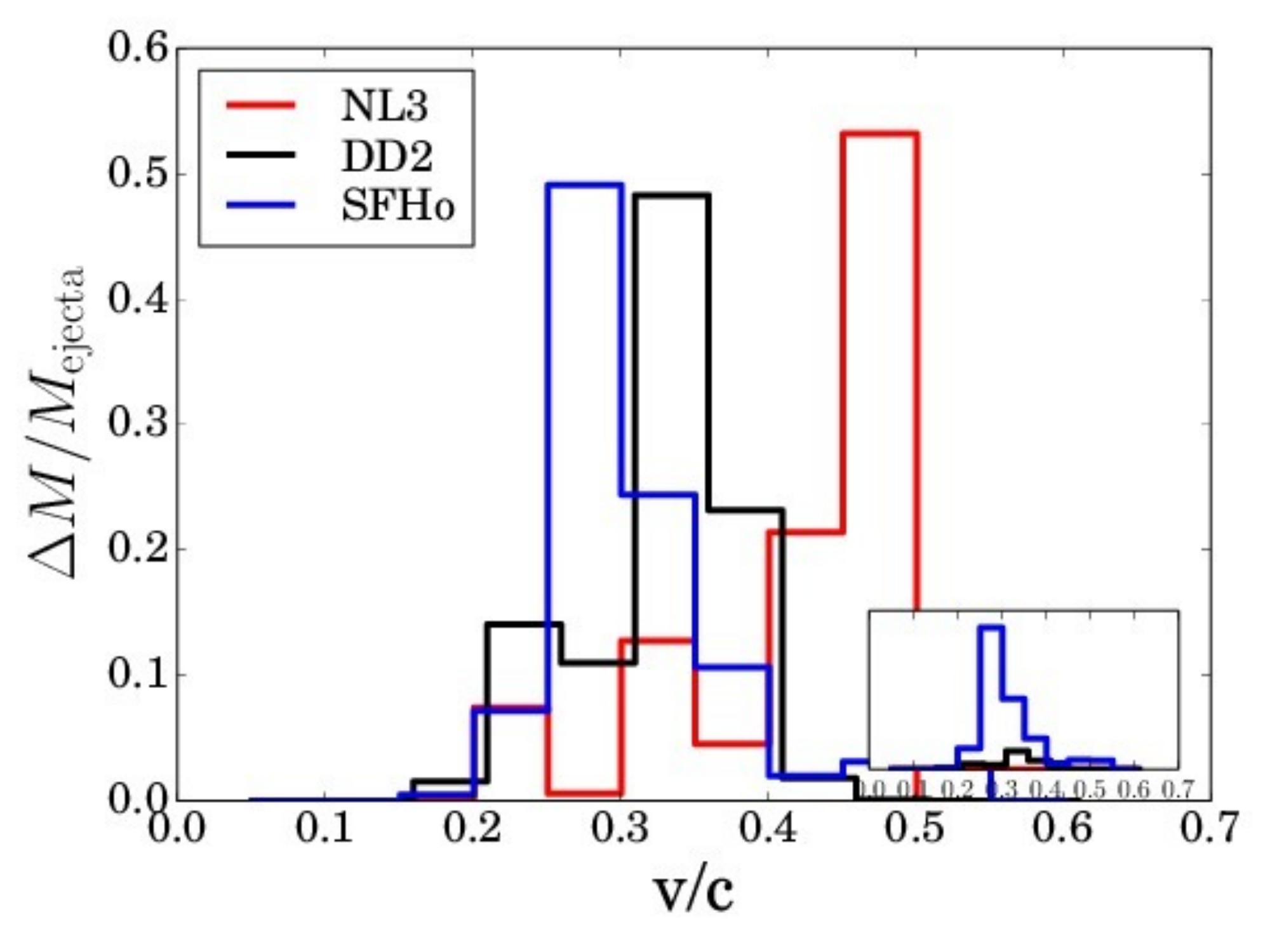}
\includegraphics[width=4.cm,angle=0]{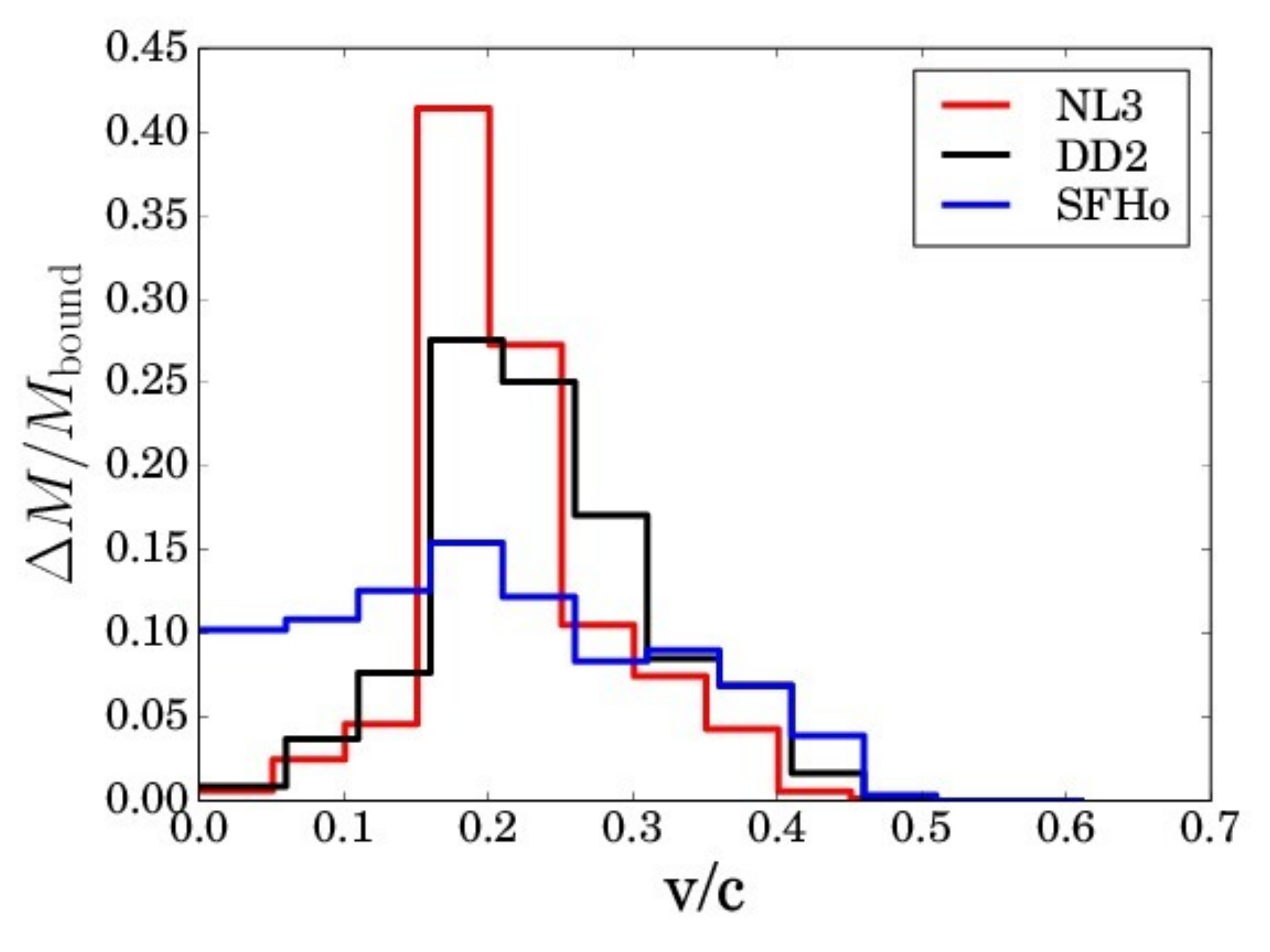}
\caption{Distributions of speeds for stellar material. (Left) The fraction of
{\bf ejected} mass binned according to velocity. The inset shows instead the absolute
mass binned in the same way. The inset demonstrates that the NL3 case has essentially
no unbound mass compared to either the DD2 or SFHo cases, and the SFHo has significantly
more unbound mass than the DD2 case.
(Right) The fraction of {\bf bound}
material with some particular velocity.
}
\label{fig:outflows_vel}
\end{figure}

\begin{figure}[h]
\centering
\includegraphics[width=4.cm,angle=0]{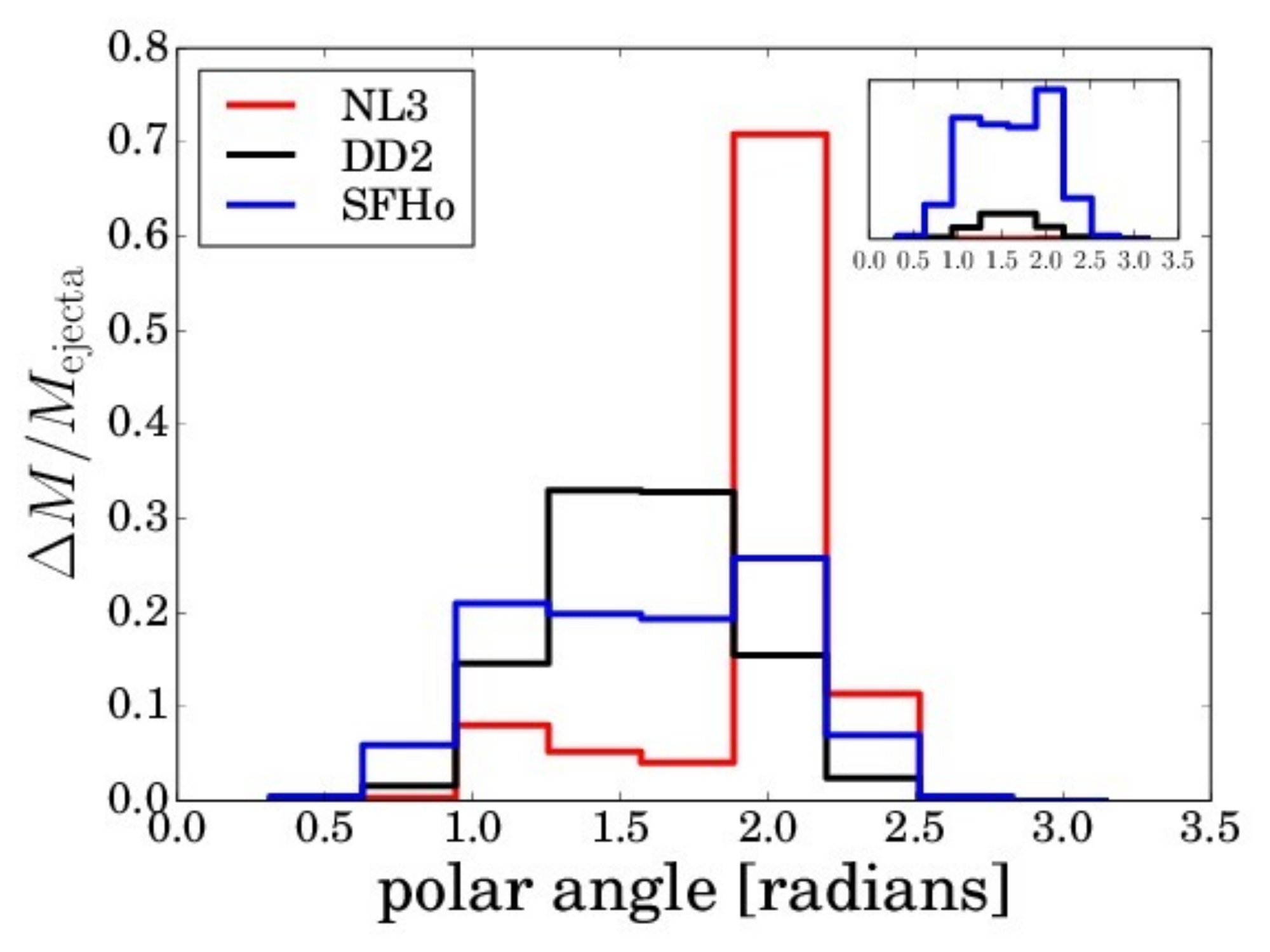}
\includegraphics[width=4.cm,angle=0]{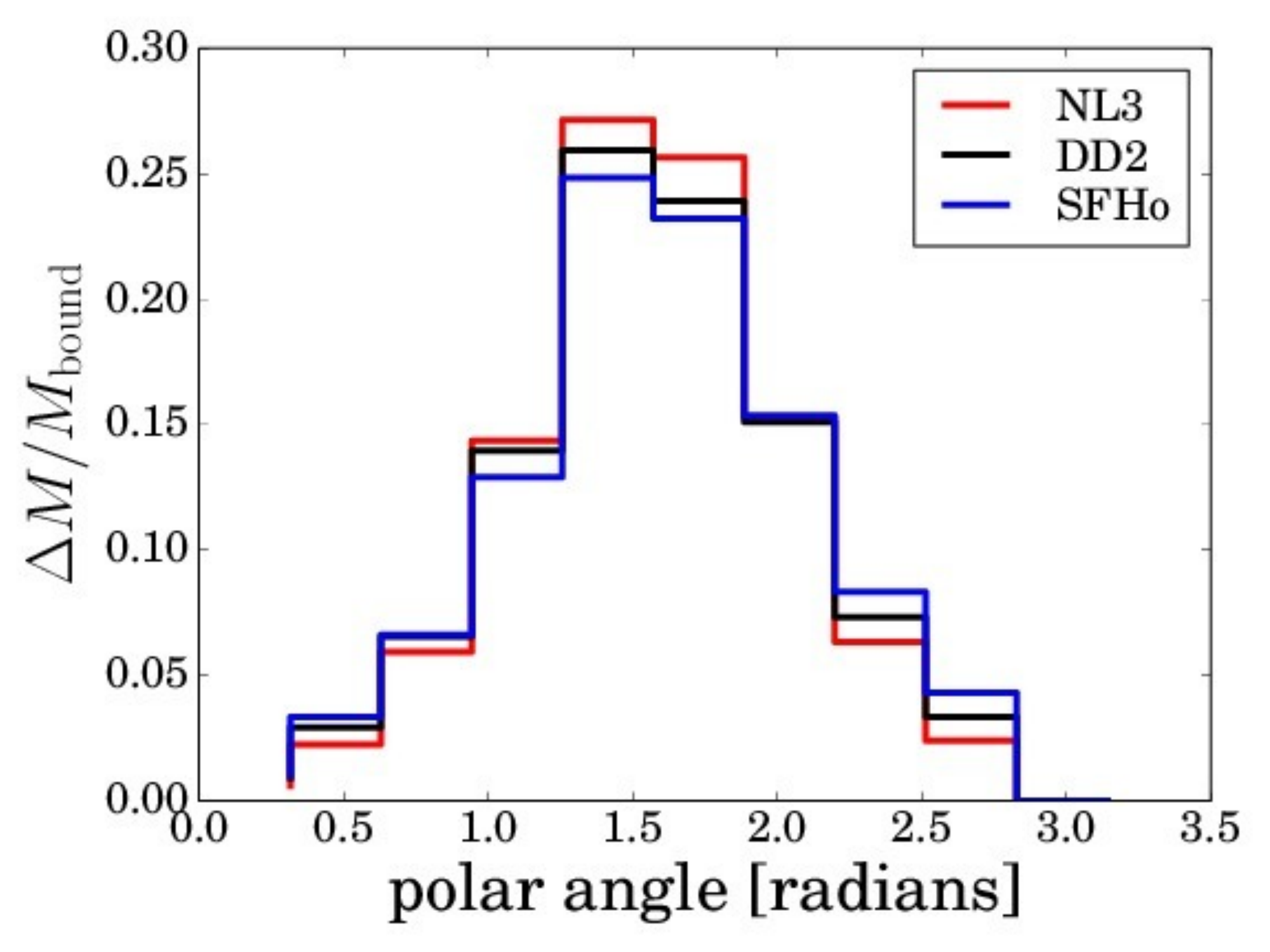}
\caption{Polar distributions of material for the different EoS cases.
Stellar material is binned
according to its polar angle. (Left) {\bf Ejected} material.
(Right) {\bf Bound} material. Both bound and ejected material is
mostly found near the orbital plane.
The asymmetry across the orbital plane for the NL3 unbound mass
is very pronounced
and probably indicative roughly of our computational error because
the amount of unbound material for that case is so small. 
} 
\label{fig:outflows_cones}
\end{figure}

\begin{figure}[h]
\centering
\includegraphics[width=4.cm,angle=0]{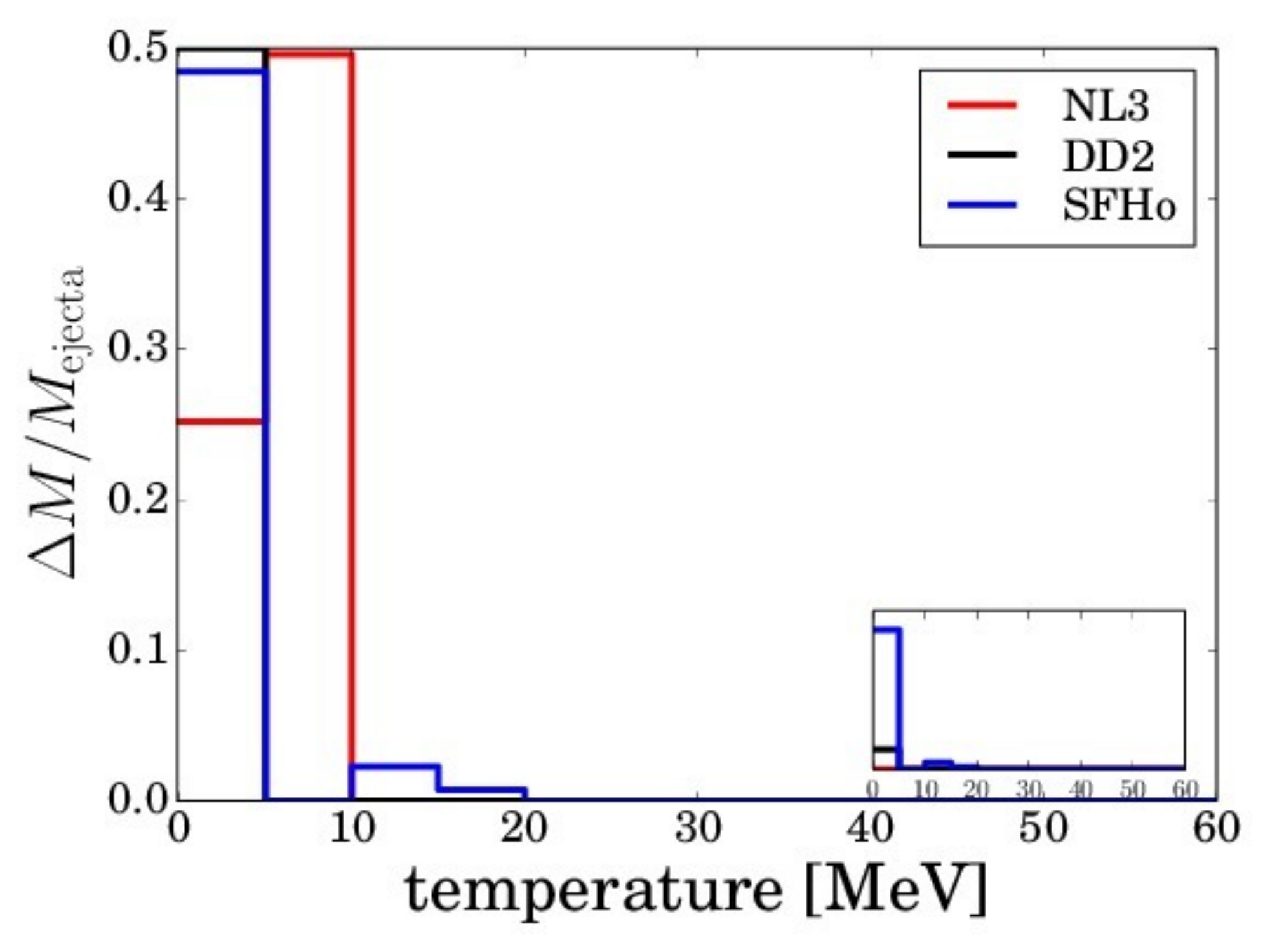}
\includegraphics[width=4.cm,angle=0]{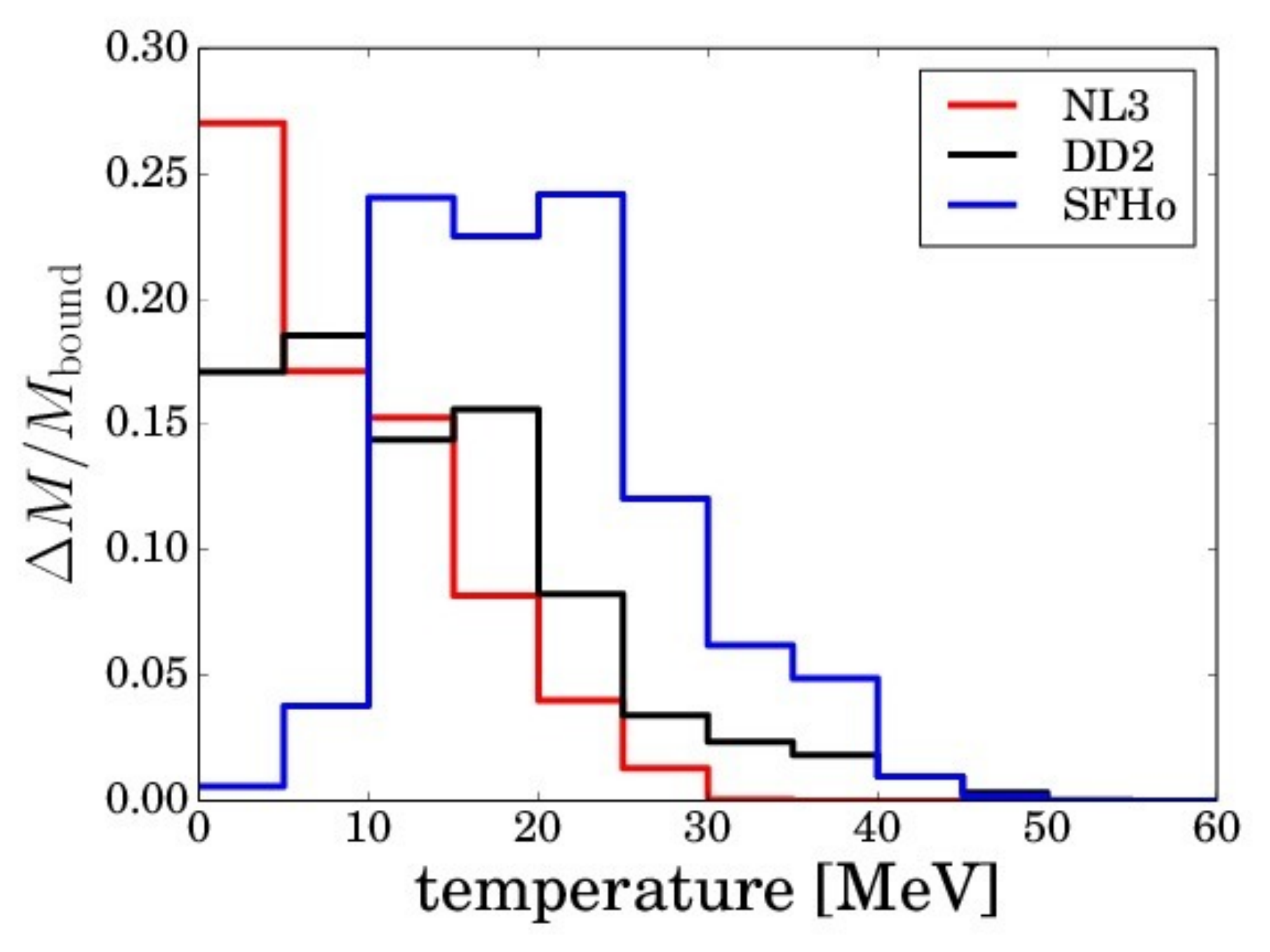}
\caption{Distributions in temperature of stellar material. (Left) Unbound
material is generally cool for all three EoS. (Right) Bound material.
The temperature of the SFHo case is noticeably higher than that of the
stiff EoS.
} 
\label{fig:outflows_temp}
\end{figure}

\begin{figure}[h]
\centering
\includegraphics[width=4.cm,angle=0]{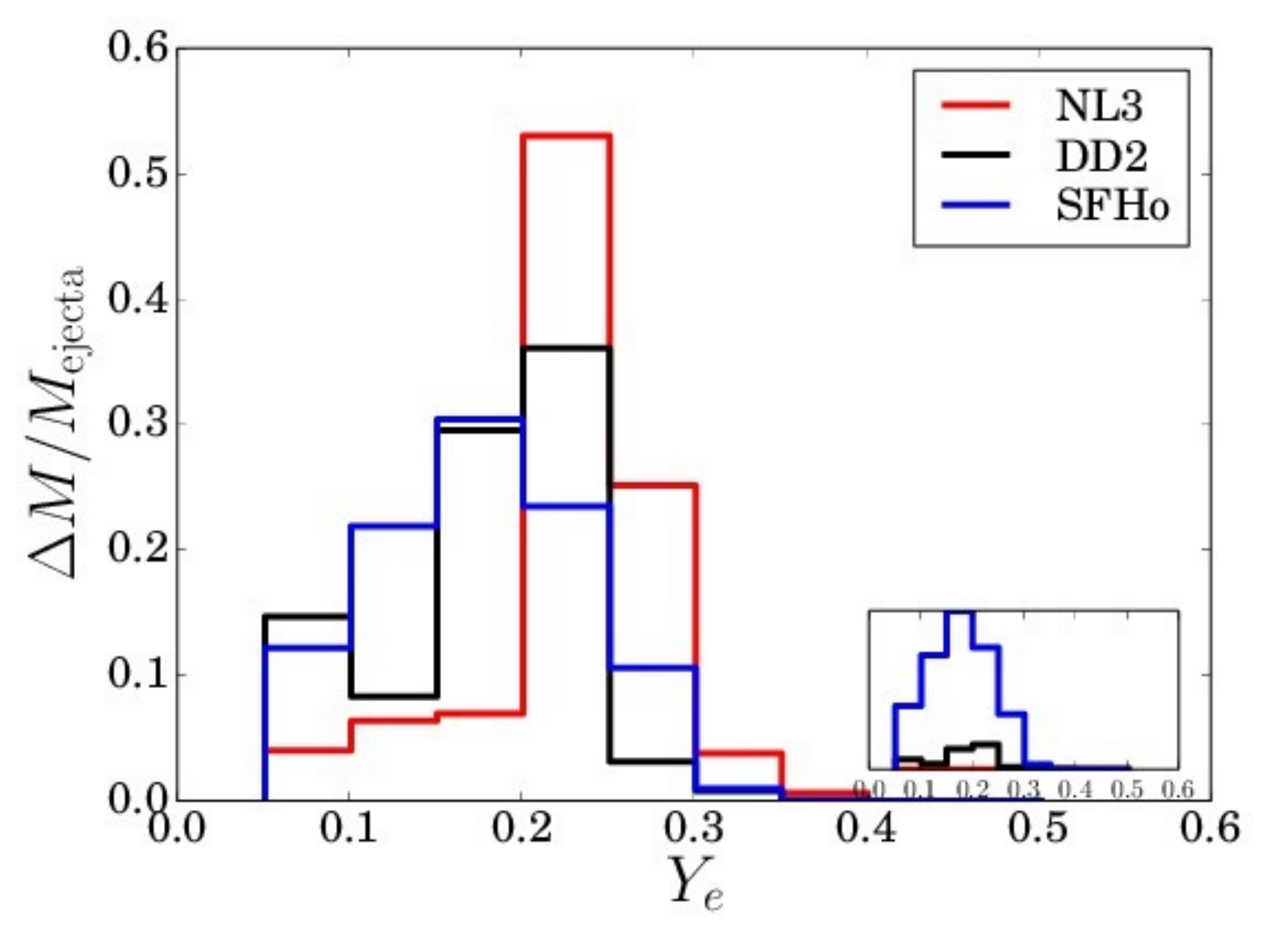}
\includegraphics[width=4.cm,angle=0]{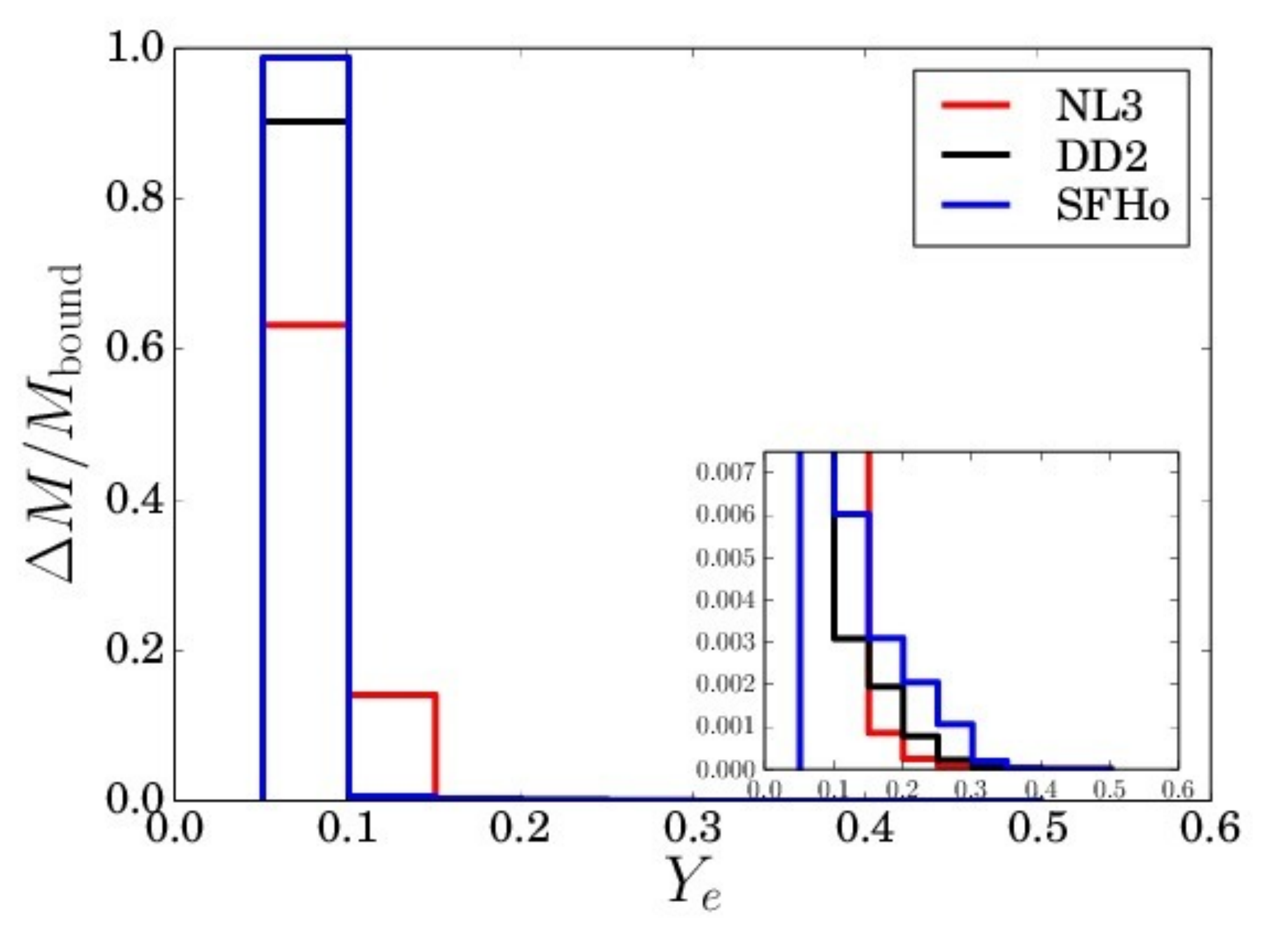}
\caption{Distributions of electron fraction for the stellar material. 
(Left) {\bf Unbound} material for the different EoS. The inset shows
the unrescaled amount of unbound material with the given electron
fraction.
Note that the outflow material is neutron rich ($Y_e < 0.3$).
(Right) {\bf Bound} material. The inset here does not show absolute
quantities, but instead shows the same information as the main
plot but with much smaller vertical scales.
} 
\label{fig:outflows_ye}
\end{figure}

To be more quantitative, we present a number of histograms that characterize the statistics of this post-merger material, measured
at roughly $t=3$ ms after the merger. We concentrate on the behavior of the matter 
at this early time to ensure that atmosphere effects do not
contaminate the extracted values of the outflows. Note however, that this implies that our quantities of unbound material might be consequently lower than the actual values, as matter can be ejected not only due to the tidal interactions, but also because of thermal pressure, which  increases significantly at the merger due to shock heating.
We present the velocity distribution in Fig.~\ref{fig:outflows_vel}, as well as 
directional dependence (as measured with respect to the direction of the orbital angular momentum) in Fig.~\ref{fig:outflows_cones}, 
the temperature in Fig.~\ref{fig:outflows_temp}, and the electron-fraction in Fig.~\ref{fig:outflows_ye}.
An important distinction is whether this material is bound to the remnant or whether it is unbound ejecta.

When looking
at the unbound material, the reader should understand that the histograms display the fractional masses of
material with some property. The huge disparity in unbound material among the EoS is therefore not immediately
apparent. Therefore, for the unbound matter we include histograms of the absolute quantities in the insets; the stiff EoS material
is barely visible.
In particular, SFHo has $0.0032 M_{\odot}$ unbound material,
the DD2 has roughly $13\%$ of this amount and the NL3 has only $0.05\%$ (these amounts are also included in Table~\ref{table:equal_mass}).
That a soft EoS yields more ejecta agrees with recent results of Ref.~\cite{2015arXiv150206660S}, and correlates with the
collision occurring deeper in the gravitational potential than for the stiffer EoS.
The observation of an apparent kilonova~\cite{2013Natur.500..547T,Berger:2013wna}, which would require significant
ejecta, is therefore suggestive of a soft-intermediate EoS, provided the binary is composed of two neutron stars
with similar masses. We will return to this point in the conclusion.

Fig.~\ref{fig:outflows_vel} illustrates the
fraction of ejected and bound mass, binned according to velocity.
While the distribution in speeds of the bound material is within a similar
range $v \in (0,0.5)\, c$, although the velocities of the ejecta are quite different given the narrower 
peak for the stiffer equation of state.
The stiffest EoS, NL3, has large velocities. It is not clear, however, if these high velocities
have any physical significance because the quantity of ejecta is so tiny.

We display the polar angular distribution of the post-merger material in
Fig.~\ref{fig:outflows_cones}. The bound material is mostly
in the equatorial plane, regardless of EoS. Unbound material is similarly
distributed mostly near the equatorial plane.  Both these observations imply the system
avoids baryon poisoning along the polar regions and so winds could be preferentially channeled
along such regions. Since such winds can also trigger r-processes though seemingly with a larger value
of $Y_e$~\cite{Kasen:2014toa,Fernandez:2014bra} and thus likely to induce signals through radioactive decay
in the optical and ultra-violet bands.

Fig.~\ref{fig:outflows_temp} contrasts the temperature  distribution among the EoS. The temperatures
of unbound material are definitively low for all three EoS. However, the bound material of the SFHo case
is generally much hotter than the other two EoS (i.e., with a distribution peaked at temperatures
in ($10$--$20$) MeV in the SFHo case as opposed to  $0$--$5$ MeV for the stiffer ones, as well as a tail with temperatures
higher by $10$MeV with respect to the tail in temperatures measured for the DD2 and NL3 cases. As mentioned, the soft EoS yields a more violent
collision with concomitant higher temperatures. The largest fraction of SFHo bound material is
close to $30$ MeV while the stiff peaks at roughly $10$--$20$ MeV.

Finally, Fig.~\ref{fig:outflows_ye} shows the electron fraction distribution and, in particular, that
the unbound material is generally quite neutron rich. The calculated values have a significant fraction
below  $Y_e \lesssim 0.22$, a rather robust condition for r-process nucleosynthesis being able to
create high mass-number elements~\cite{Kasen:2014toa,lippunerprivate}. Consequently, an electromagnetic
signal peaking in the infrared would be expected in such cases.

\subsection{Neutrino Emission}

\begin{figure}[h]
\centering
\includegraphics[width=9.2cm,angle=0]{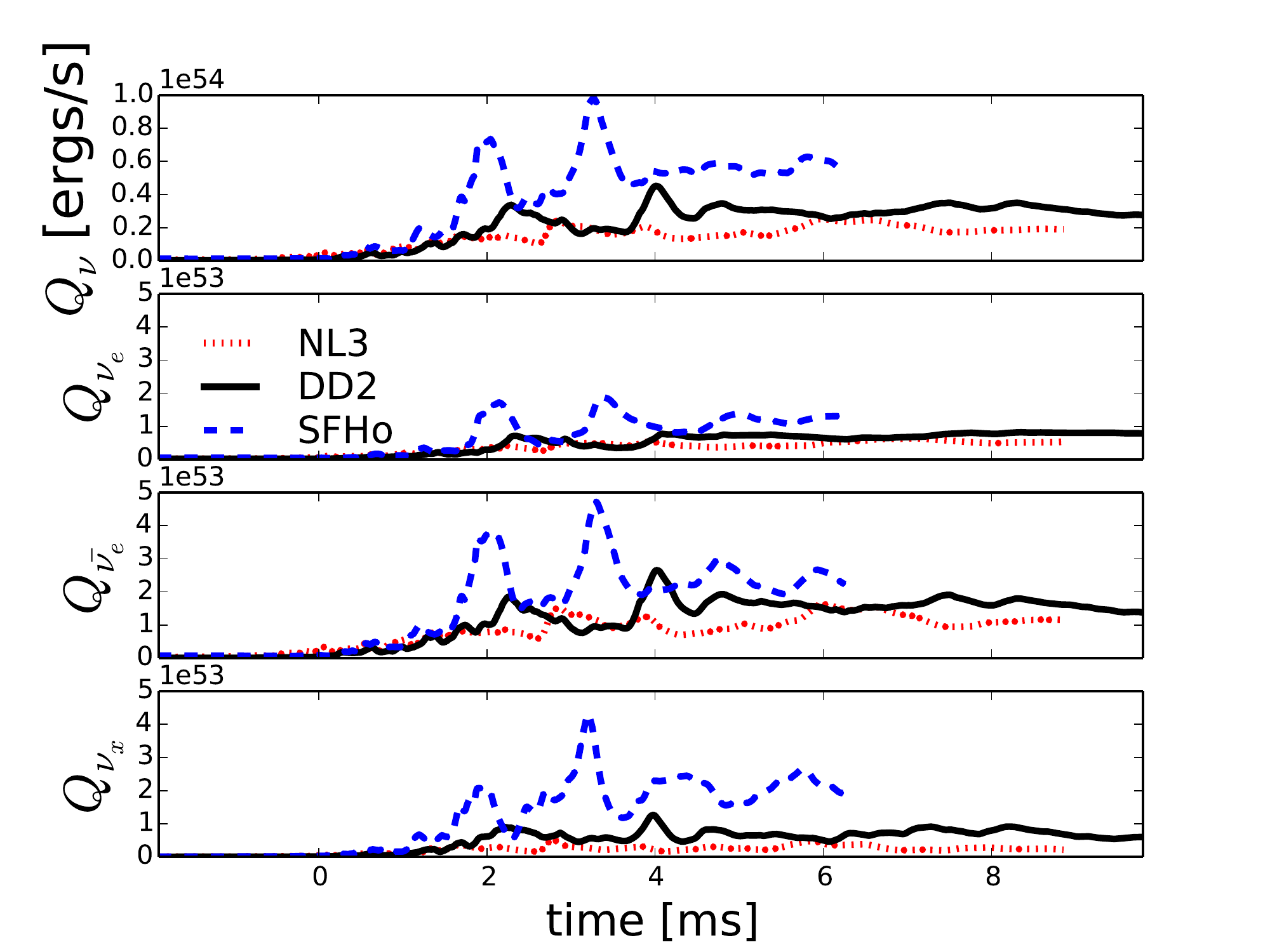}
\caption{ Neutrino luminosities for the different EoS.
} 
\label{fig:luminosity_neutrino}
\end{figure}

We compute neutrino cooling and the corresponding luminosity through a
leakage scheme as reported in~\cite{Neilsen:2014hha}.
Fig.~\ref{fig:luminosity_neutrino} shows the luminosities
for each neutrino species and the total neutrino luminosity as a
function of time for each EoS. Following along the trend of the
previous results which showed the largest gravitational wave emission
and largest ejecta mass for the SFHo EoS, we also see the largest
neutrino luminosity for this EoS. This reflects the fact that the
temperature and decompressed mass achieved in this case is the
highest. Irrespective of the particular EoS, it is evident that the
luminosities for both electron neutrinos and electron antineutrinos
become roughly comparable while the heavy-lepton neutrino types are
less luminous (we note that the heavy lepton luminosity shown here is
actually the contribution for all four heavy-lepton species).  This
dominance has already been observed in other binary neutron star
mergers~\cite{Sekiguchi:2011zd, Sekiguchi:2012uc, Neilsen:2014hha} and
also neutron star--black hole mergers~\cite{Deaton:2013sla}, and in
simulations with more complete neutrino
transport~\cite{2015arXiv150206660S, Foucart:15}. This phenomena is
due to neutron rich material being shock heated and
decompressed~\cite{1997A&A...319..122R}.  This neutron rich, low
density, hot material is initially far below the new
$\beta$-equilibrium value of $Y_e$ and will preferentially emit
electron antineutrinos until equilibrium is reestablished.

Given the high neutrino luminosities achieved during the merger it is
interesting to study their possible detection in future and current
facilities. For the purposes of detection on Earth, we are most
interested in the spectrum and luminosity of the electron antineutrinos,
as these neutrinos will produce the largest signal in
Earth-based water Cherenkov detectors, such as the current
Super-Kamiokande or IceCube detectors, or the future Hyper-Kamiokande
detector.  It is true that neutrino flavor oscillations can occur due to one or
more of the following: (1) neutrino-neutrino interactions, 
(2) neutrino-matter interactions, or (3) oscillations in vacuum.
 In these cases, some fraction of the electron
antineutrino spectrum will oscillate into heavy lepton antineutrinos
and vice versa. The precise influence of neutrino oscillations on the
electron antineutrino spectrum observed in Earth based detectors is
unknown, but depends sensitively on the unknown neutrino mass hierarchy and
the neutrino and matter conditions in the merger remnant~\cite{malkus:14}. For this
reason, in our analysis we neglect oscillations and assume the
electron antineutrino signal at Earth is the emitted 
signal from our merger calculation with a gravitational redshift. 
We note that Ref.~\cite{Foucart:15} 
recently showed  that the electron antineutrino
luminosity predicted from their leakage scheme, which is similar in
design to ours, matches the prediction from a more complete neutrino
transport calculation to better than $\sim$~30\,\%. Furthermore, the
neutrino luminosities shown in Fig.~\ref{fig:luminosity_neutrino} agree
well with the recent results of \cite{2015arXiv150206660S}, who study
similar EoS and initial conditions. The leakage scheme also predicts a
lepton loss rate that, in theory, provides an average energy for the
emitted neutrinos.  However, this average energy prediction is
typically too high because it includes contributions from very high
energy neutrinos leaking out from optically thick zones.  In reality
these neutrinos thermalize as they diffuse out.  A more representative
average energy is one determined by the thermodynamic conditions near
the neutrinosphere, the location where the neutrinos leave
thermodynamic equilibrium with the matter.

To make a more accurate determination of this neutrino average energy,
we use the formalism described in Ref.~\cite{caballero:09}.  We
chose to examine the disk as seen by an observer looking down from
$z=+\infty$, i.e. looking face-on towards the disk.  To locate the
neutrinosphere of the remnant disks for this direction, rays are
traced down onto the disk from $z = +\infty$ and halted when the
integrated optical depth reaches 2/3. We repeat this for many $(x,y)$
pairs.  We ignore general relativistic effects on the ray
propagation. The stopping location sets the
neutrinosphere surface, $\mathcal{S}$. 
During the integration, at a given spatial location,
we assume the opacity is the average over a Fermi-Dirac, zero chemical
potential neutrino number distribution with a temperature equal to the
local matter temperature. As contributions to the opacity we include:
(1) elastic scattering of all neutrino species on neutrons, protons, and
electrons; (2) neutrino capture on protons (neutrons) for electron
antineutrinos (electron neutrinos); and (3) neutrino-antineutrino
annihilation. As we include scattering cross sections, our neutrino
surfaces correspond to the surface of last scattering, not necessarily
the surface where the neutrinos thermodynamically decouple.  The
differences between these two surfaces is small for electron-type
neutrinos and antineutrinos because the cross sections are dominated
by the charged current processes, but this difference is larger for
the heavy lepton neutrinos which do not have such interactions in BNS
merger remnants. That being said, to determine the average energy of
neutrinos emitted in this direction, for a particular neutrino
species, we average over $\mathcal{S}$, and compute
\begin{equation}
\langle E_{\nu_i} \rangle = \int_\mathcal{S} dA \int dE\,  E \phi_{\nu_i}(E,\mathcal{S})
\bigg/ \int_\mathcal{S} dA \int dE \phi_{\nu_i}(E,\mathcal{S})\,,\label{eq:averageenergy}
\end{equation}
where the integral is carried out over the neutrinosphere surface $\mathcal{S}$, determined
by the rays; the Fermi-Dirac flux, $\phi(E,\mathcal{S})$, is given by
\begin{equation}
\phi(E,\mathcal{S} )=\frac{c}{2\pi^2(\hbar c)^3}E^2f_{\rm FD},
\end{equation}
where $f_{\rm FD}$ is the Fermi-Dirac
distribution with a temperature equal to the local matter temperature
on the surface $\mathcal{S}$. The neutrino energy will be
gravitationally redshifted as neutrinos stream away from the merger remnant.
We estimate the neutrino energy at infinity by transforming the emitted quantities using
the invariance of phase-space density,
\begin{equation}
\frac{1}{c^2}\frac{dN}{d^3xd^3p}=\frac{f_{\rm FD}}{h^3c^2}= \frac{I}{E^3},
\end{equation}
where $I$ is the specific intensity, and
\begin{equation}
 \frac{dN}{d^3xd^3p}=\frac{c^2}{E^2}\frac{dN}{dAdtdEd\Omega}.
\end{equation}
To proceed, let us denote the quantities measured by an observer with a ``tilde'', 
(i.e. the observed energy is $\tilde E$, the area differential is $d\tilde A$, etc.) 
while the analogous emitted quantities are $E$ and $dA$, etc. Now, by invariance
of $I/E^3$ we have
\begin{equation}
  \frac{1}{c^2}\frac{dN}{d^3\tilde xd^3\tilde p}=\frac{1}{c^2}\frac{dN}{d^3xd^3p}.
\end{equation}
Focusing on the left hand side,
\begin{equation}
  \frac{1}{c^2}\frac{dN}{d^3\tilde xd^3\tilde p}=\frac{1}{\tilde E^2}\frac{dN}{d\tilde E d \tilde A d\tilde t d \tilde \Omega}=\frac{f_{\rm FD}}{h^3c^2}
\end{equation}
we obtain,
\begin{equation}
\frac{dN}{d\tilde t}=\int\frac{f_{\rm FD}}{h^3c^2} \tilde E^2d\tilde E d \tilde A d \tilde\Omega. \label{dndt}
\end{equation}
We can now rewrite Eq.~(\ref{dndt}) 
in terms of the emitted quantities given that $\tilde E =\sqrt{g_{00}} E$, assuming
that the emission is isotropic, and that distances are stretched by a factor of $(1-r_s/r)^{-1/2}=g^{-1/2}_{00}$
(utilizing the Schwarzschild metric for simplicity). Thus, Eq.~(\ref{dndt}) can be re-expressed as
\begin{equation}
\frac{dN}{d\tilde t}=4\pi \int \frac{f_{\rm FD}}{h^3c^2} g_{00}^{3/2} E^2dE \frac{d A}{g_{00}}= 4\pi \int \frac{f_{\rm FD}}{h^3c^2} g_{00}^{1/2} E^2dE d A,
\end{equation}
and thus
\begin{equation}
\frac{dN}{d\tilde t}= \int g^{1/2}_{00} \phi(E,\mathcal{S})dE\,dA.
\end{equation}
We can proceed in a similar way for $\tilde E dN/d \tilde t=d\tilde E/d \tilde t$, and find,
\begin{equation}
 \frac{d \tilde E}{d\tilde t}=\int \frac{4\pi}{h^3c^2} g_{00}\frac{E^3}{e^{E/T}+1}dE d A=\int g_{00} \phi(E,\mathcal{S})EdEdA. \label{dedt}
\end{equation}
This agrees with the well known result that the luminosity at infinity is $L_\infty= g_{00}L$ in the
case where the neutrino surface is at a constant redshift.

Eq.~(\ref{dedt}) contains $g_{00}$ which must be evaluated at the neutrino surface. Then the
average observed energy is $\langle E_{\nu_i} \rangle=(d\tilde E/d\tilde t)/ (dN/d \tilde t)$.
For simplicity, when evaluating the numerical value of
$g_{00}$, we employ the Schwarzschild metric (in Schwarzschild coordinates) with a central
black hole mass of $2.7\,M_\odot$.  Inside 25\,km ($\sim$ 3
Schwarzschild radii), to avoid spuriously overestimating the amount of redshift, 
we fix the value of the metric to the value at 25\,km, i.e.  $|g_{00}|(r<25$\,km$) \sim 0.68$. (For the
purpose of our estimates, this simple evaluation does not depend sensitively on these details however.)
 We present the average energies computed this way for several times
and EoS in Table~\ref{tab:energies}.

\begin{table}
\begin{center}
\begin{tabular}{c|cccc}
 Time&$\langle E_{\bar\nu_e}\rangle$ &$\langle E_{\nu_e}\rangle$ &$L_{\bar{\nu}_e}$&$R_\nu$\\
    $[$ms$]$  & $[$MeV$]$&$[$MeV$]$ &$[10^{53}$ erg/s$]$&$[$\#/ms$]$\\
  \hline
\phantom{0}2.5 (NL3)\phantom{o} &18.5 (22.4)& 15.2 (18.3) &0.71&18.1\\
\phantom{0}3.0 (DD2)\phantom{o} &18.3 (22.1)& 14.6 (17.4)& 1.1\phantom{0}&28.2\\
\phantom{0}3.2 (SFHo) &24.6 (29.7)& 23.5 (28.3) &3.5\phantom{0}&120.8\\
\hline
\phantom{0}8.4 (NL3)\phantom{o}& 13.4 (15.6) & 9.8 (11.3)&1.1\phantom{0}&20.7\\
\phantom{0}7.9 (DD2)\phantom{o}& 13.2 (16.1) & 10.2 (12.4)&1.6\phantom{0}&29.6\\
\hline
\end{tabular} 
\caption{Observed (emitted, with no gravitational redshift taken into
  account in parenthesis) neutrino average energies for electron
  antineutrinos and electron neutrinos, the electron antineutrino
  luminosity predicted from the leakage scheme (from Fig.~\ref{fig:luminosity_neutrino}), and the estimated
  instantaneous detection rate in Super-Kamiokande for a merger at
  10\,kpc from Earth via Eq.~\ref{eq:detectionrate} for two different
  times and three EoS. In the text, we give estimates of the integrated
  event rate up to the 6\,ms after merger.}
\label{tab:energies}
\end{center}
\end{table}

\begin{figure}
\begin{center}
\includegraphics[angle=-90, trim=4cm 0cm 0cm 2cm, clip=true,width=\columnwidth]{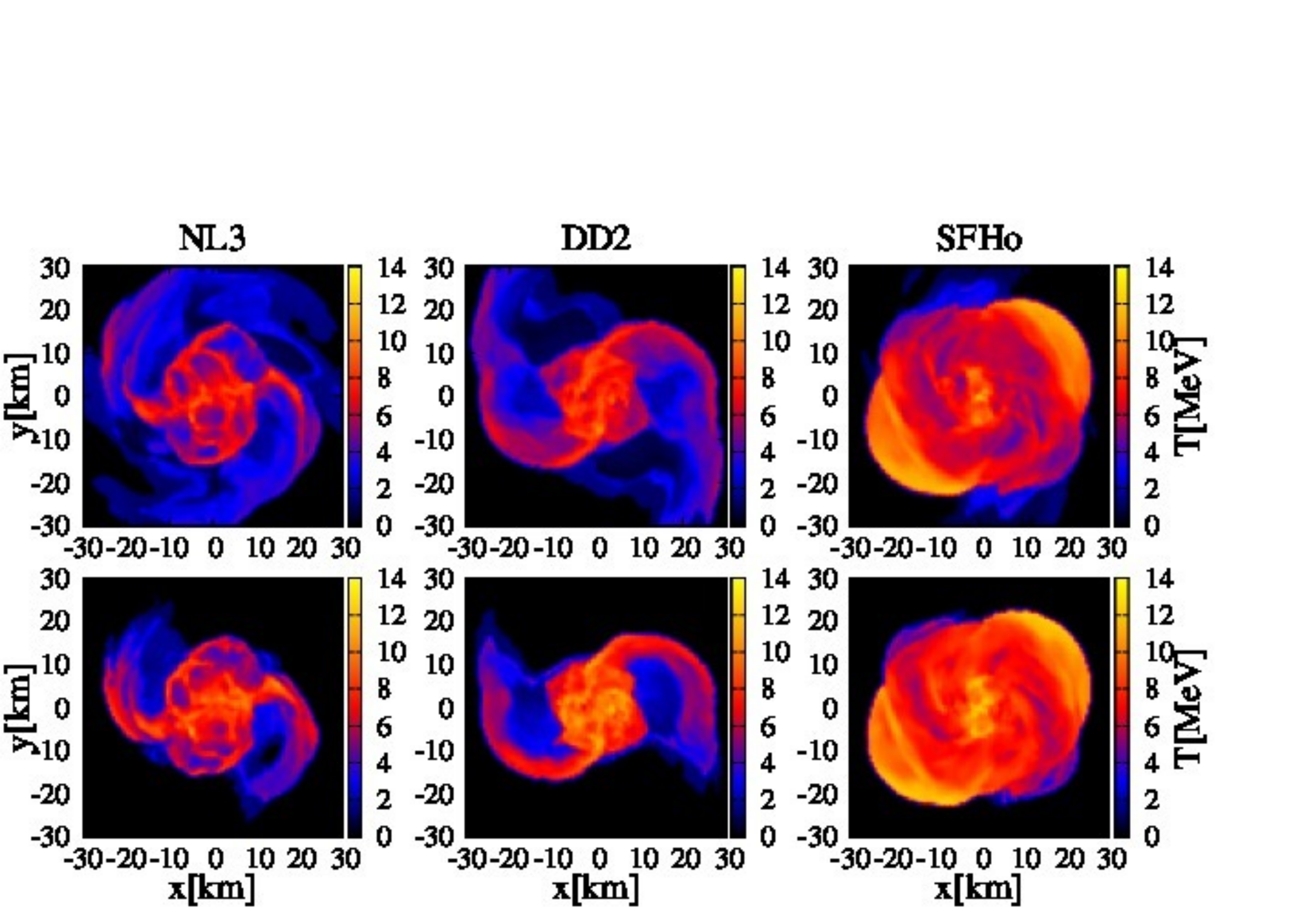}
\caption{Color map of the temperature at the scattering
  neutrinospheres at $\sim 2.5$--$3$\,ms after merger for the NL3 EoS
  (left), DD2 EoS (center) and SFHo EoS (right).  The top panels show
  the electron neutrinosphere, and the bottom panels show the electron
  antineutrinosphere.}
\label{fig:nuEOS}
\end{center}
\end{figure}

In Fig.~\ref{fig:nuEOS}, we show snapshots of the spatial distribution
of the temperature at the neutrinosphere at $\approx 2.5$--3\,ms after
merger for the three EoS and for each electron-type neutrino species. As the matter is
primarily neutron rich, the electron neutrino neutrinosphere has a
larger extent than the electron antineutrinosphere.
The SFHo merger produces the most decompressed material and ejecta
and the highest temperatures; thus powering the largest neutrino luminosity. 
This EoS also has larger and
generally hotter neutrinospheres for all species. For the NL3 and DD2
EoS, the neutrinosphere is generally confined to the MNS except for
the region around the tidal tails.  These tails are cold and have
little mass (see \S III.B).

Armed with the neutrino luminosities and average energies, we can 
estimate the detection rate at Earth-based neutrino
detectors. Given the uncertainties introduced by the leakage scheme,
we make the following assumptions:  (1) We assume that the
luminosities shown in Fig.~\ref{fig:luminosity_neutrino} are radiated
isotropically from the merger. Observers situated along the merger
axis would observe a larger than average neutrino luminosity, while
observers situated on, or near, the equator would observe smaller
average luminosities~\cite{Perego:14}.  Therefore, the following predictions are
the average expected rate over many sky orientations.  (2)  We
assume the spectrum of neutrinos follows a zero chemical potential,
Fermi-Dirac spectrum with a neutrino temperature that gives the
average energy computed above
($\langle E_{\nu_i} \rangle \sim 3.15T_{\nu_i}$). (3) We assume
that if an interaction occurs in the volume of the detector, it will
be registered as an event. This may not be true for the lowest energy
neutrinos, but such events should be few in number due to the low
cross section. (4) Finally, as mentioned above, we neglect
neutrino oscillations which may alter the electron antineutrino signal
at Earth. Given these assumptions, the detection rate in a
water-Cherenkov detector is then roughly given as
\begin{equation}
R_\nu \sim N_T \times n_\nu \times \frac{\int
\mathcal{F}^2_{T_\nu} \sigma_\mathrm{IBD} dE} {\int \mathcal{F}^2_{T_\nu} dE}\,,
\end{equation}
where $N_T$ is the number of target protons, which for the 32kT
Super-Kamiokande is equal to
$2/18\times 32\times10^9\, \mathrm{g} / m_\mathrm{amu} \sim
2.1\times10^{33}$,
$n_{\nu} = L/\langle E_\nu \rangle / (4\pi D^2)$ is the total
number-flux of neutrinos a distance $D$ from the source, and
$\int \mathcal{F}^2_{T_\nu} \sigma_\mathrm{IBD} dE / \int
\mathcal{F}^2_{T_\nu} dE$
is the spectrum weighted average cross section with
$\mathcal{F}^2_{T} = E^2/[\exp(E/T) + 1]$ equal to the zero-chemical
potential Fermi-Dirac neutrino number distribution and where
$\sigma_\mathrm{IBD}$ is the inverse-beta decay cross section, which
can be approximated as
$\sigma_\mathrm{IBD} \sim \sigma_0 / (4 m_e^2) (1+3g_A^2)
E_{\bar{\nu}_e}^2$
with $\sigma_0 = 1.761\times10^{-44}\mathrm{cm}^2$, $m_e$ =
0.511\,MeV, $g_A$ = -1.254.  All together, the detection rate is
\begin{equation}
R_\nu \sim \frac{21.1}{\mathrm{ms}}\left[\frac{32\mathrm{kT}}{M_\mathrm{water}}\right]\left[\frac{L_\nu}{10^{53}\,
  \mathrm{erg}/\mathrm{s}}\right]\left[\frac{\langle E_\nu \rangle}{15\,\mathrm{MeV}}\right]\left[\frac{10\,\mathrm{kpc}}{D}\right]^2\,.\label{eq:detectionrate}
\end{equation}

 
We have used Eq.~\ref{eq:detectionrate} to determine detection rates
for the times and EoS listed in Table~\ref{tab:energies}.  We take the
electron antineutrino luminosities from
Fig.~\ref{fig:luminosity_neutrino} for the corresponding time and EoS
(which we include in Table~\ref{tab:energies} for completeness).  The
resulting instantaneous detection rates for a merger located at
10\,kpc from Earth in a detector like Super-Kamiokande are listed in
the last column of Table~\ref{tab:energies}.  To estimate the total
number of neutrinos that can be detected, we take a fiducial length of
time of 6\,ms and integrate over Eq.~\ref{eq:detectionrate} using the
electron antineutrino luminosities shown in
Fig.~\ref{fig:luminosity_neutrino} and assuming a constant
$\langle E_{\bar{\nu}_e} \rangle$ equal to the value near 3\,ms in
Table~\ref{tab:energies}.  For the NL3, DD2, and SFHo EoS, we calculate
$\approx $135, $\approx $176, $\approx $405 events, respectively, in the first
6\,ms after merger in a Super-Kamiokande like detector and a merger
event at $\approx $10\,kpc. The merger with the SFHo EoS predicts much
higher rates because both the luminosity and the average energy of the
neutrinos is larger due to the higher matter temperatures achieved in
the merger. If the resulting MNS collapses to a black hole (as in the
SFHo case studied here), this could lead to a sharp reduction in the
neutrino luminosity (e.g. Ref.~\cite{2015arXiv150206660S}), otherwise,
we would expect a slower decline in the neutrino luminosity as the MNS
and disk cool. Ultimately many more events could be detected in the
MNS/disk cooling phase, and both the total number, and signal duration
could strongly depend on the EoS~\cite{Just:2015}.

One can ask how far would we be able to detect a BNS merger in
neutrinos.  As the distance of the merger increases, the rate of
detection decreases quadratically, as seen in
Eq.~(\ref{eq:detectionrate}).  For mergers similar to the ones studied
here, but located in the Andromeda galaxy $\sim$800\, kpc from Earth rather
than $10\,$kpc, the detection rates are $\sim$6,400 times
smaller. Consequently, we would not expect any neutrinos in a detector
like Super-Kamiokande. Detections with Super-Kamiokande would be
limited to the Milky Way and surrounding satellite galaxies. However,
with Hyper-Kamiokande, a future water-Cherenkov detector with a
proposed volume $\sim$20 times that of Super-Kamiokande, the expected number
of neutrinos in the first $\sim$10\,ms would be $\mathcal{O}(1)$ and maybe a
few over the entire lifetime of the disk, thus detectable especially
in light of further timing input provided by gravitational waves.

\subsection{Magnetic Effects}

We have also investigated the possible impact of magnetization within a timescale of $\approx 15$ms from
the onset of the merger. To this end, we consider for simplicity the intermediate case (DD2) and endow the
stars with a magnetic dipole of maximum strength $10^{13}$ G. 
The dynamics of the merger and early post-merger present
several instabilities and processes capable of significantly amplifying the 
magnetization ~\cite{Price:2006fi,Anderson:2008zp,Obergaulinger:2010gf,Kiuchi:2014hja}. 
Indeed, recent local simulations have shown that some of these instabilities can drive the magnetic energy density up to a level
approaching equipartition with the kinetic energy of the turbulent fluid on a timescale given by the turn-over time of the resulting
eddies~\cite{2010A&A...515A..30O,2013ApJ...769L..29Z}.
However, these process occur on very fine scales that require numerical resolution (and computational resources) generally beyond
the reach of codes such as ours.
Furthermore, with grid-based codes fully capturing these dynamics require resolutions of the order of 
$\lesssim 0.1 $m.

Instead of such exorbitant resolutions, we 
introduce an ``effective driver'' mechanism (similar to those recently used in other 
works~\cite{2014arXiv1410.0013G,2015MNRAS.447...49S}) to account for this
amplification 
by including a term in the 
evolution equation for the magnetic field
of the form 
\begin{equation}
  \partial_t (\sqrt{\gamma} B^i) = ... + \alpha \sqrt{\gamma} \xi B^i ~.
\end{equation}
Here we have defined the amplification factor as 
\begin{equation}
\xi \equiv \xi_0 \Theta(\rho [\omega^z - \omega^z_{\rm thresh}] ) g(t) B_k v^k  \, ,
\end{equation}
with $\Theta(x)$ the Heaviside function so that the amplification factor vanishes
 when the z-component of the fluid vorticity, $\omega^z$, is below some threshold $\omega^z_{\rm thresh}$. This threshold
is chosen to only activate this term for the eddies with wavelengths shorter than a third of the MNS radius.
(This avoids, in particular, adding an additional increase to the magnetic field due to the star's differential rotation.)
The factor $B_k v^k$ restricts the amplification only to the components along the velocity lines that stretch the magnetic field. 
The function $g(t)$ restricts when this sub-grid model is active as described below.
We note that the amplification introduced by this term is not explicitly divergence free. However, we monitor the
divergence of the magnetic field, and find that, while it increases at merger, the rate of increase relative
to the rate of increase of the field strength itself is comparable to that seen when the sub-grid model is
inactive. 

\begin{figure}[h]
\centering
\includegraphics[width=8.0cm,angle=0]{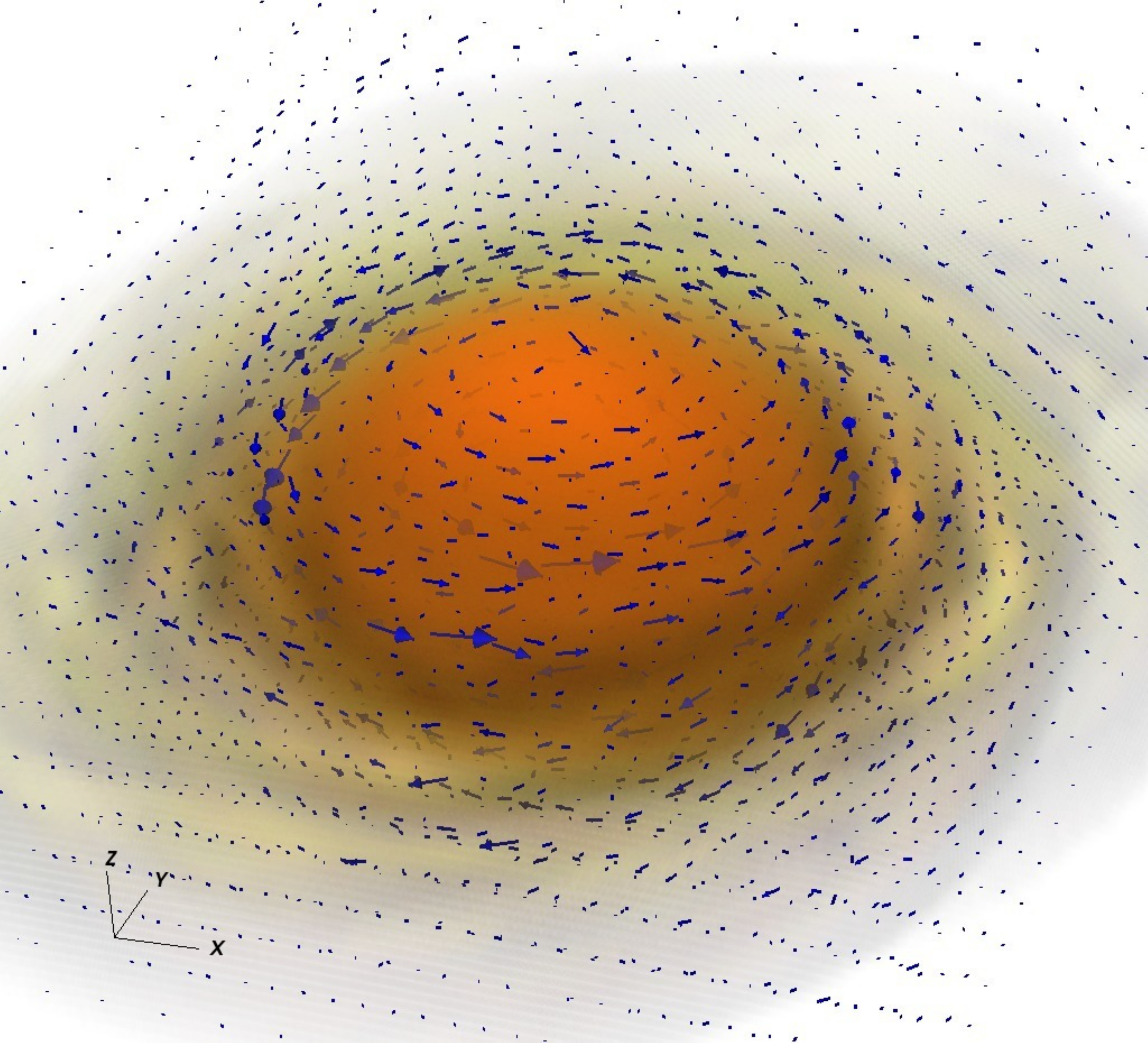}
\caption{ Density (colormap) and magnetic field vectors (blue arrows) for the high magnetization case at $t \approx 15 \rm{ms}$ after the
merger, when the system has settled into a quasi-stationary configuration.
The vectors show that the magnetic field is mostly toroidal.
} 
\label{fig:magnetization_3D}
\end{figure}

This added term must be adjusted to match the expectations established by local simulations~\cite{2013ApJ...769L..29Z}.
The term should result in the amplification of the magnetic field during merger until the associated
magnetic energy density reaches strengths close to equipartition. In particular, small-scale dynamos act to amplify
the magnetic field until the magnetic energy density reaches roughly $\approx 60\%$ of the local 
turbulent kinetic energy density~\cite{2013ApJ...769L..29Z}. 
We therefore allow this sub-grid model to act for a few milliseconds until the maximum magnetic field reaches $10^{16}$--$10^{17}$~G
(see for instance Ref.~\cite{2013ApJ...775...87D} for a formal calculation of this turbulent kinetic energy). The bulk dynamics, combined with this sub-grid model,
give rise to a MNS with a strong and largely toroidal magnetic field structure. It is worth mentioning that we have tried an amplification factor without the factor $B_k v^k$ and obtained similar results. The ensuing topology is illustrated in Fig.~\ref{fig:magnetization_3D}.

\begin{figure}[h]
\centering
\includegraphics[width=8.5cm,angle=0]{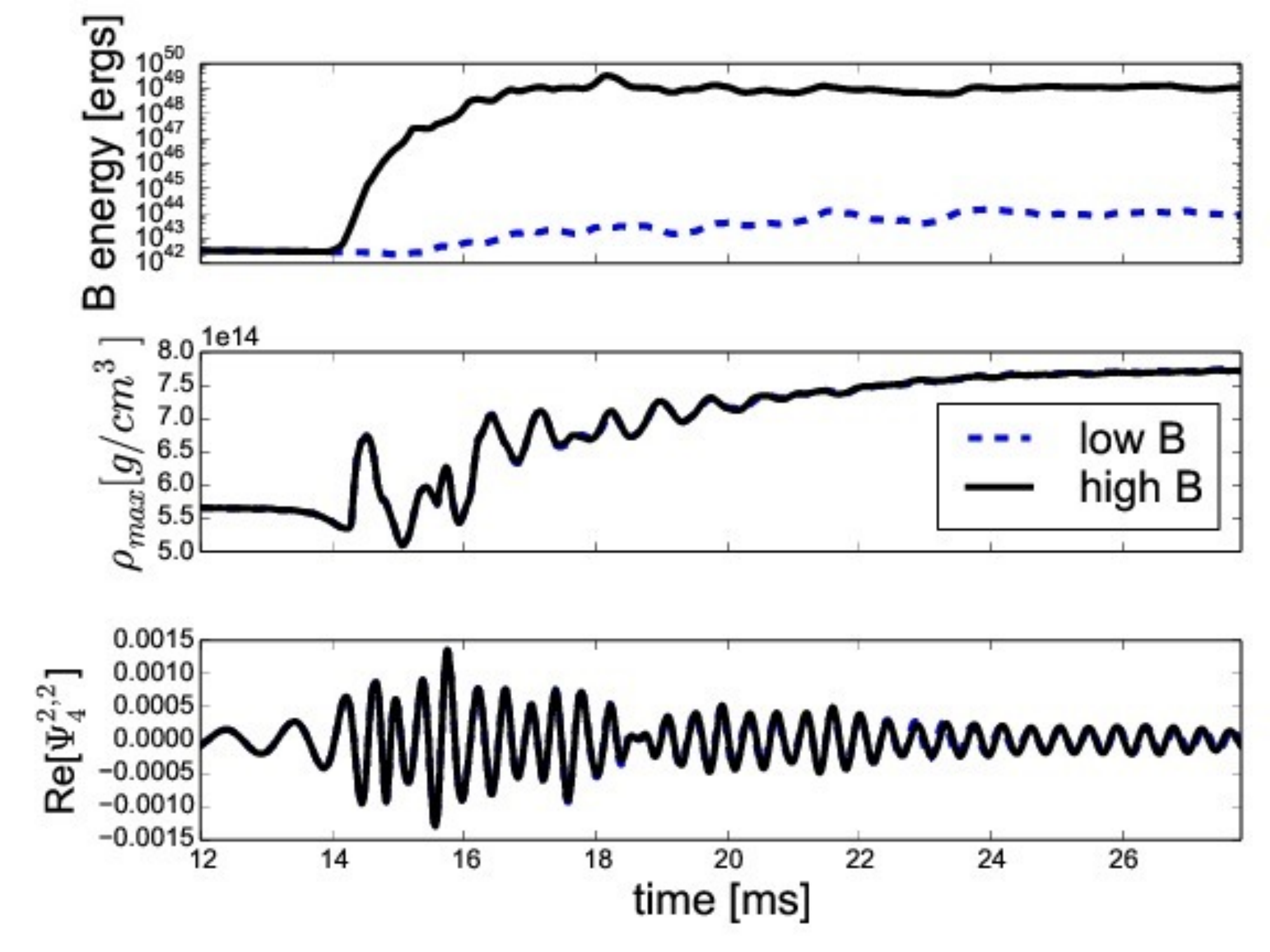}
\caption{ Comparison of low and high magnetized cases for the DD2
binary. (top) The total magnetic energy for the domain. 
(middle) The maximum density for the two cases.
(bottom) The primary mode of the GW signal.
The merger takes place at $t\approx 14.3$ ms and during this merger the magnetic
field grows significantly.
Even for the
high magnetization case in which $B_{\rm max} \approx 10^{17}$G, the stellar
dynamics remains largely unaffected as shown in the bottom two panels
primarily because the magnetic field is mostly toroidal.
Other quantities, such as the temperature and the neutrino radiation rate,
also display small differences between low and high magnetizations.
} 
\label{fig:magnetization_1dcomparison}
\end{figure}

We display the magnetic energy of
the star in the top frame of Fig.~\ref{fig:magnetization_1dcomparison}. 
We call the evolution with no sub-grid model our {\em low magnetization} case and compare it to the case
with the sub-grid model active, our {\em high magnetization}. Also included in Fig.~\ref{fig:magnetization_1dcomparison}
are plots of the maximum density (middle panel) and the resulting GW signal (bottom panel). These figures reveal essentially
no difference between the two magnetization cases within this timescale, despite the high magnetization case reaching
almost $10^{17}$ G. Nevertheless, subtle differences arise with finer-scale phenomena. 

\begin{figure}[h]
\centering
\includegraphics[width=8.8cm,angle=0]{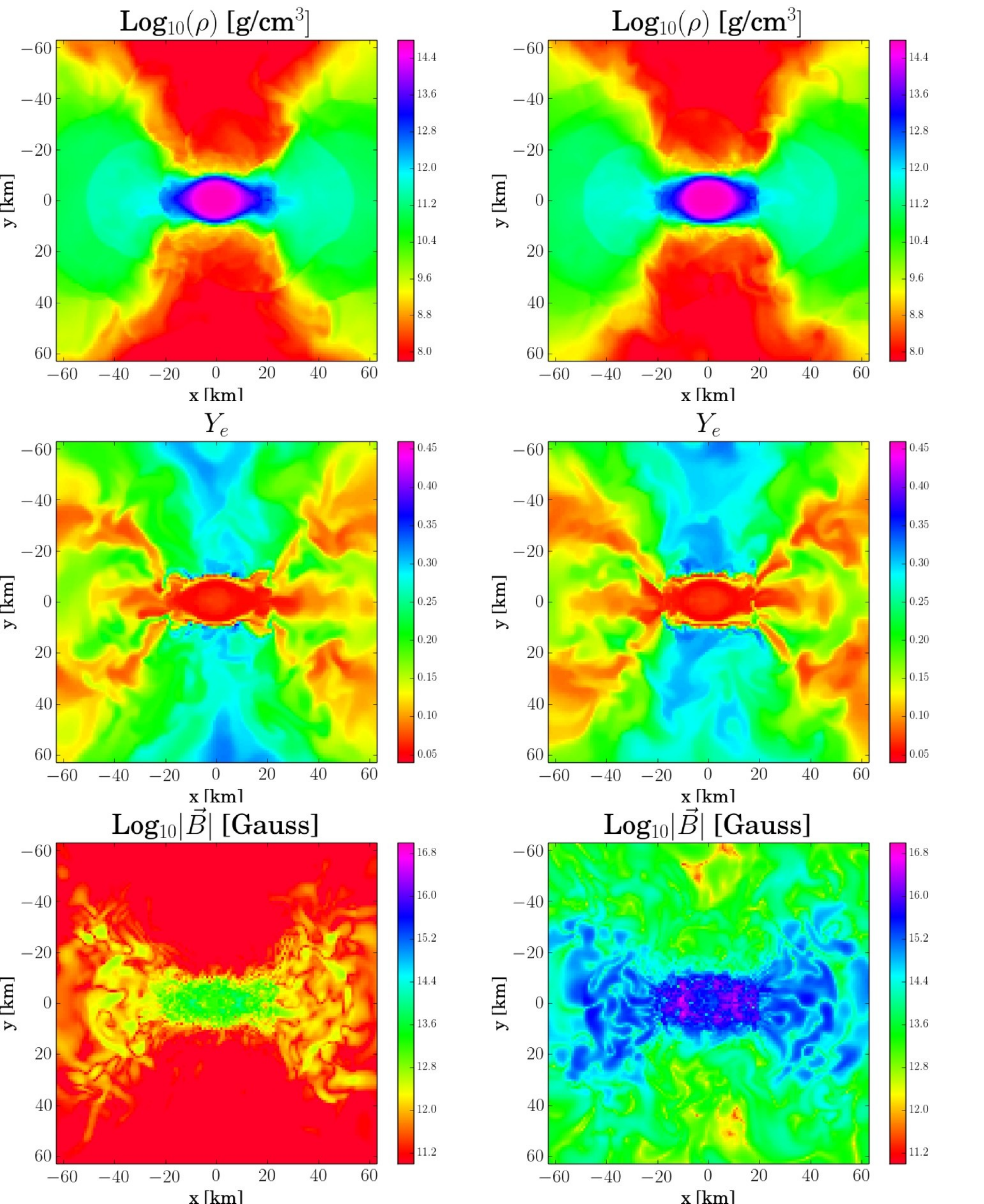}
\caption{ Snapshots of density, electron fraction, and magnetic field strength in the meridional plane $x=0$
at $t \approx 9 \rm{ms}$ after the
merger, when the system has settled into a quasi-stationary configuration. 
The low magnetization case is displayed on the left and the high magnetization case is shown on the right.
} 
\label{fig:magnetization_2dcomparisonx}
\end{figure}

\begin{figure}[h]
\centering
\includegraphics[width=4.cm,angle=0]{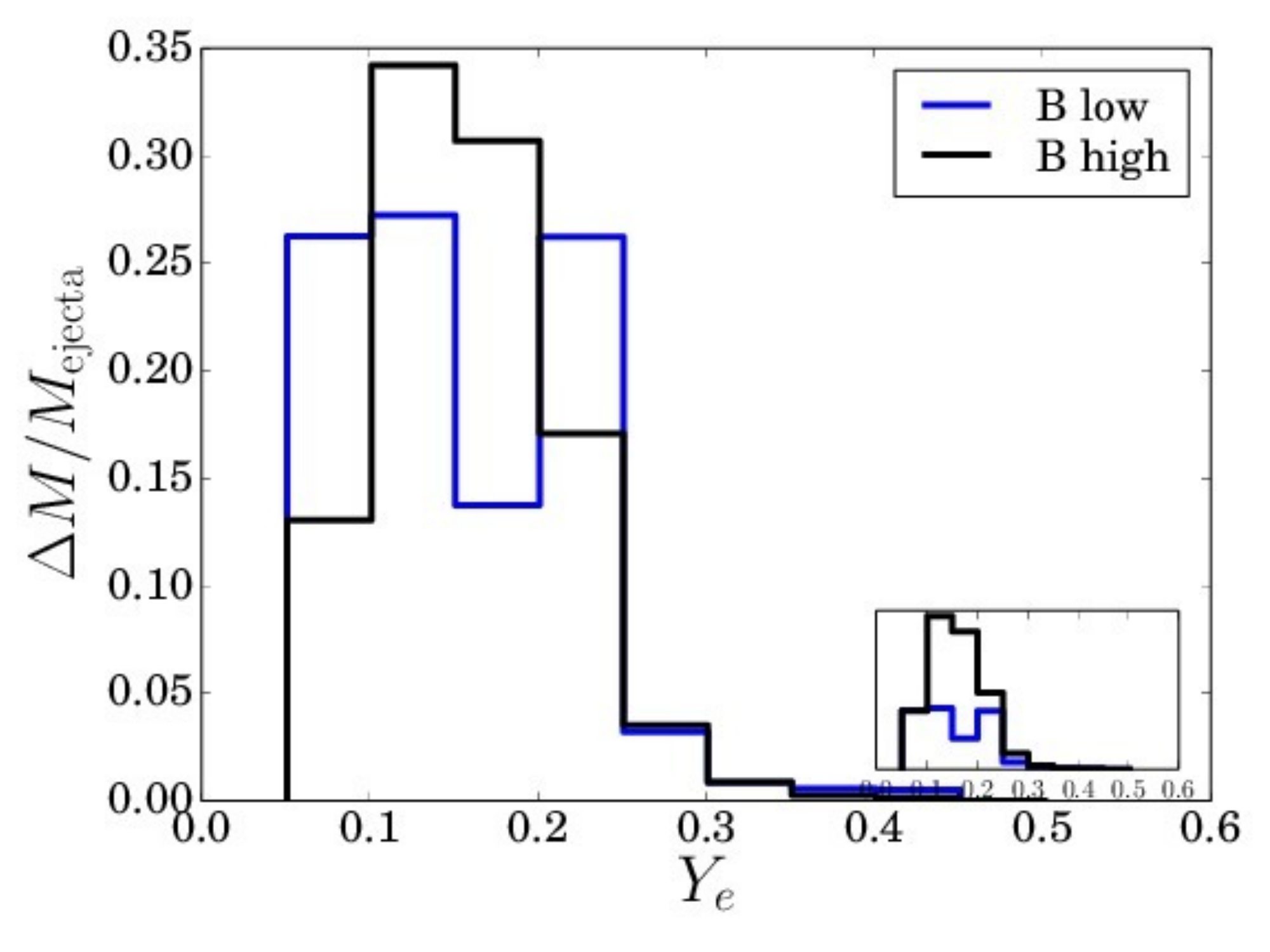}
\includegraphics[width=4.cm,angle=0]{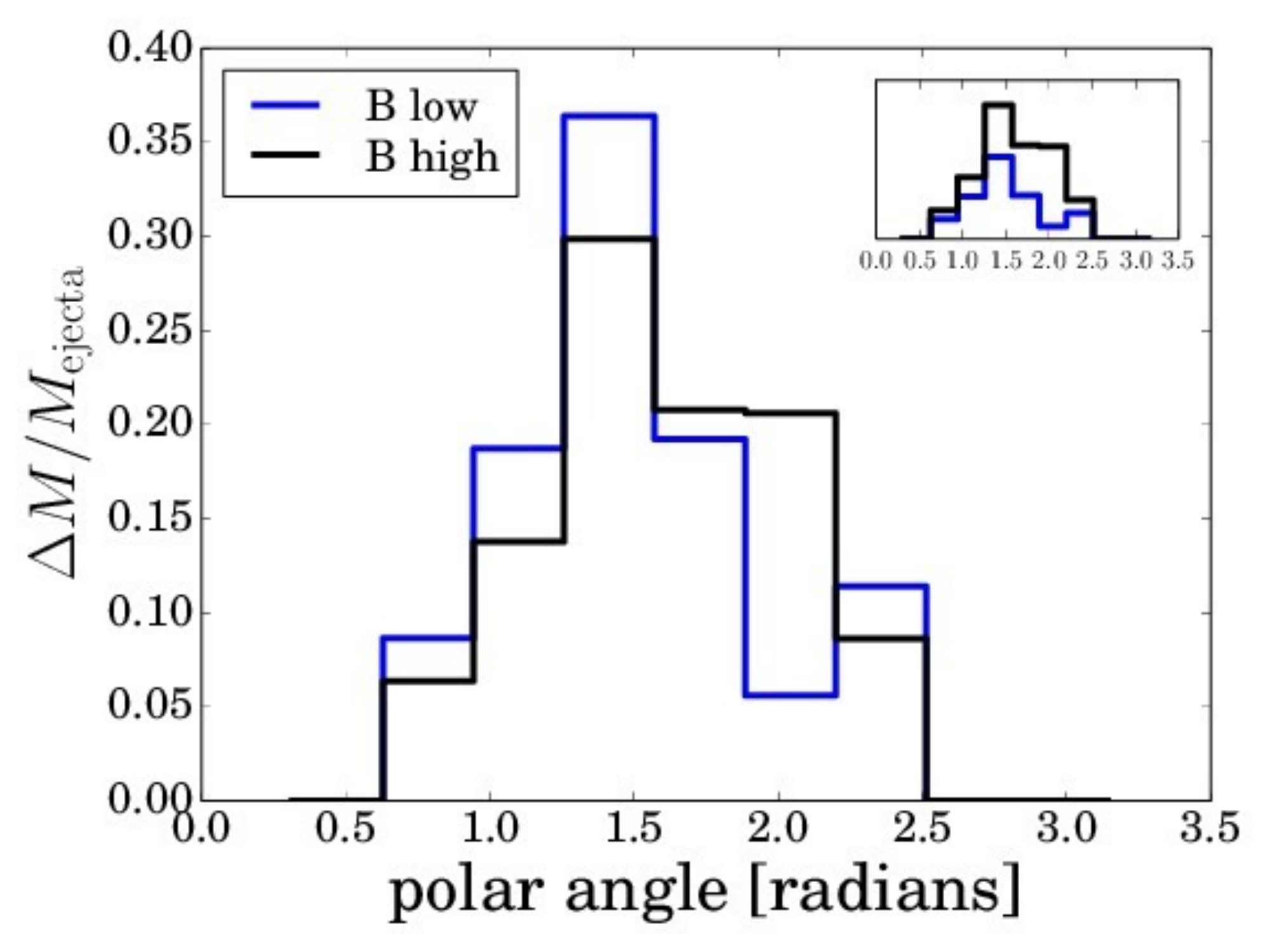}
\caption{Distributions of $Y_e$ and angular direction for unbound stellar material for the low/high magnetized cases at $t\approx 9 \rm{ms}$ after
the merger. 
}
\label{fig:outflows_YA_Bfield}
\end{figure}

We display the density, electron fraction, and magnetic field strength for both the low and high magnetization cases 
on a meridional plane in Fig.~\ref{fig:magnetization_2dcomparisonx}. A difference in electron fraction appears in
the polar regions, but other differences are quite difficult to distinguish. And so we compute the distributions
of different quantities and display them in histograms. In particular, Fig.~\ref{fig:outflows_YA_Bfield} displays
the electron fraction for unbound material. While the low case has a flat distribution in $Y_e$, the high
magnetization case peaks at roughly $Y_e \approx 0.15$. Note also that the high case ejects roughly twice
the material of the low case, presumably due to the additional magnetic pressure.
The angular distribution of the unbound matter is also shown in Fig.~\ref{fig:outflows_YA_Bfield}. The angular distribution
suggests that the extra material unbound by the high magnetic field occurs primarily along equatorial directions.

\begin{figure}[h]
\centering
\includegraphics[width=4.cm,angle=0]{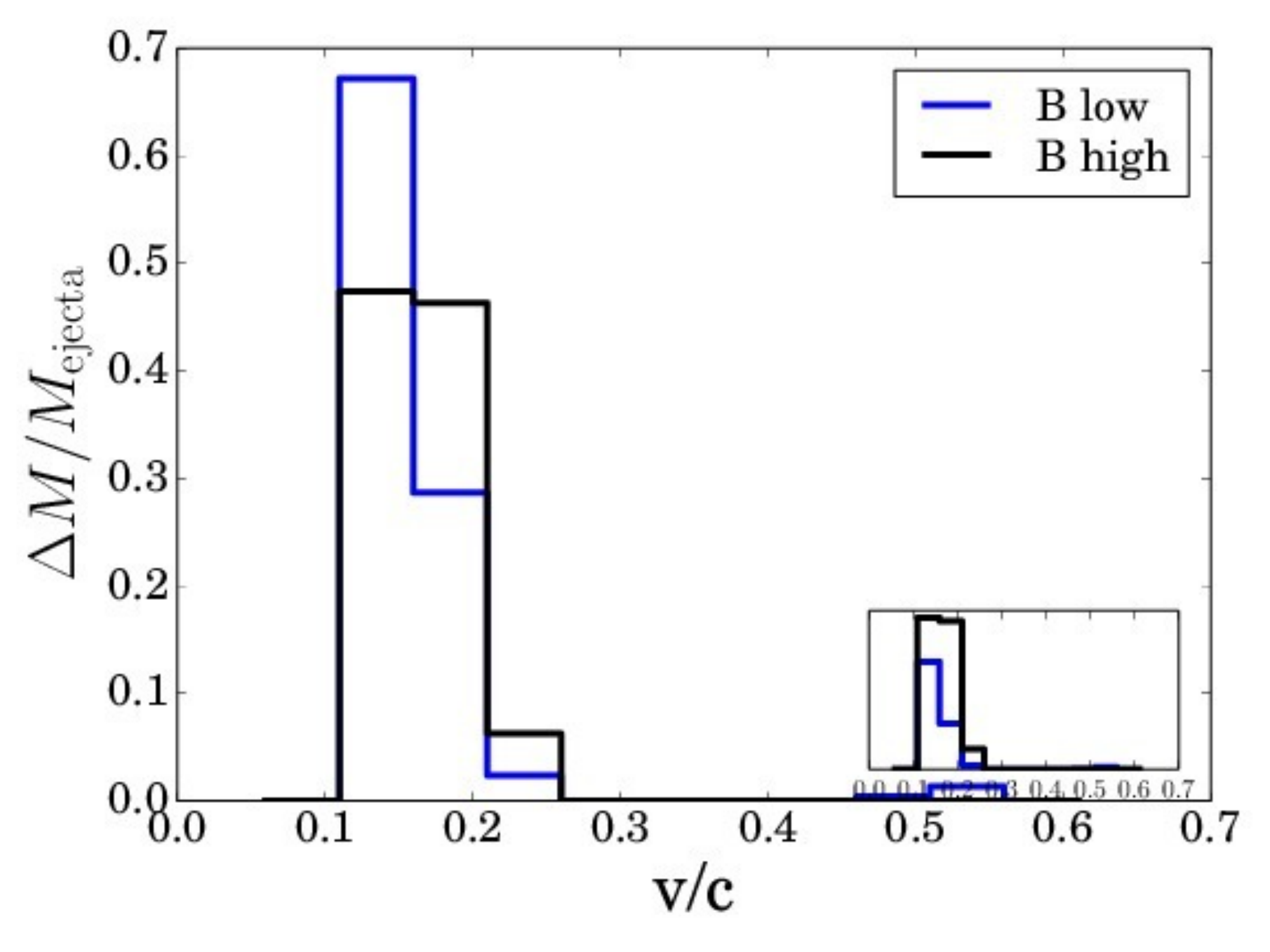}
\includegraphics[width=4.cm,angle=0]{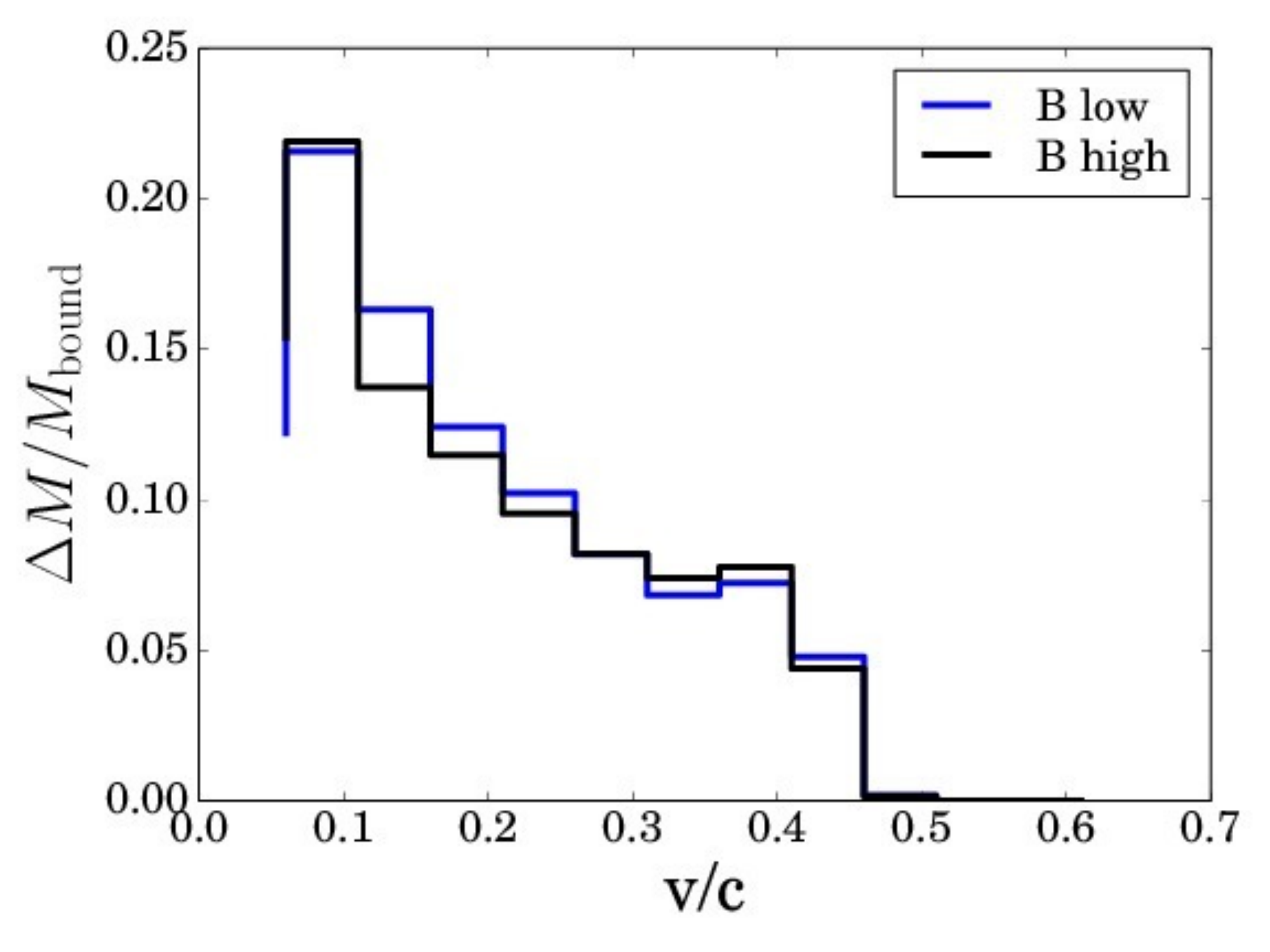}
\caption{Distributions of speeds for stellar material for the low/high magnetized cases at $t\approx 9 \rm{ms}$ after the merger.
}
\label{fig:outflows_vel_Bfield}
\end{figure}

The velocity distributions for both bound and unbound matter are shown in
 Fig.~\ref{fig:outflows_vel_Bfield}. 
The bound matter distributions for the two cases are largely the same, but the unbound matter has a slightly higher
average (but not peak) velocity for the high magnetization case.

\begin{figure}[h]
\centering
\includegraphics[width=6.5cm,angle=0]{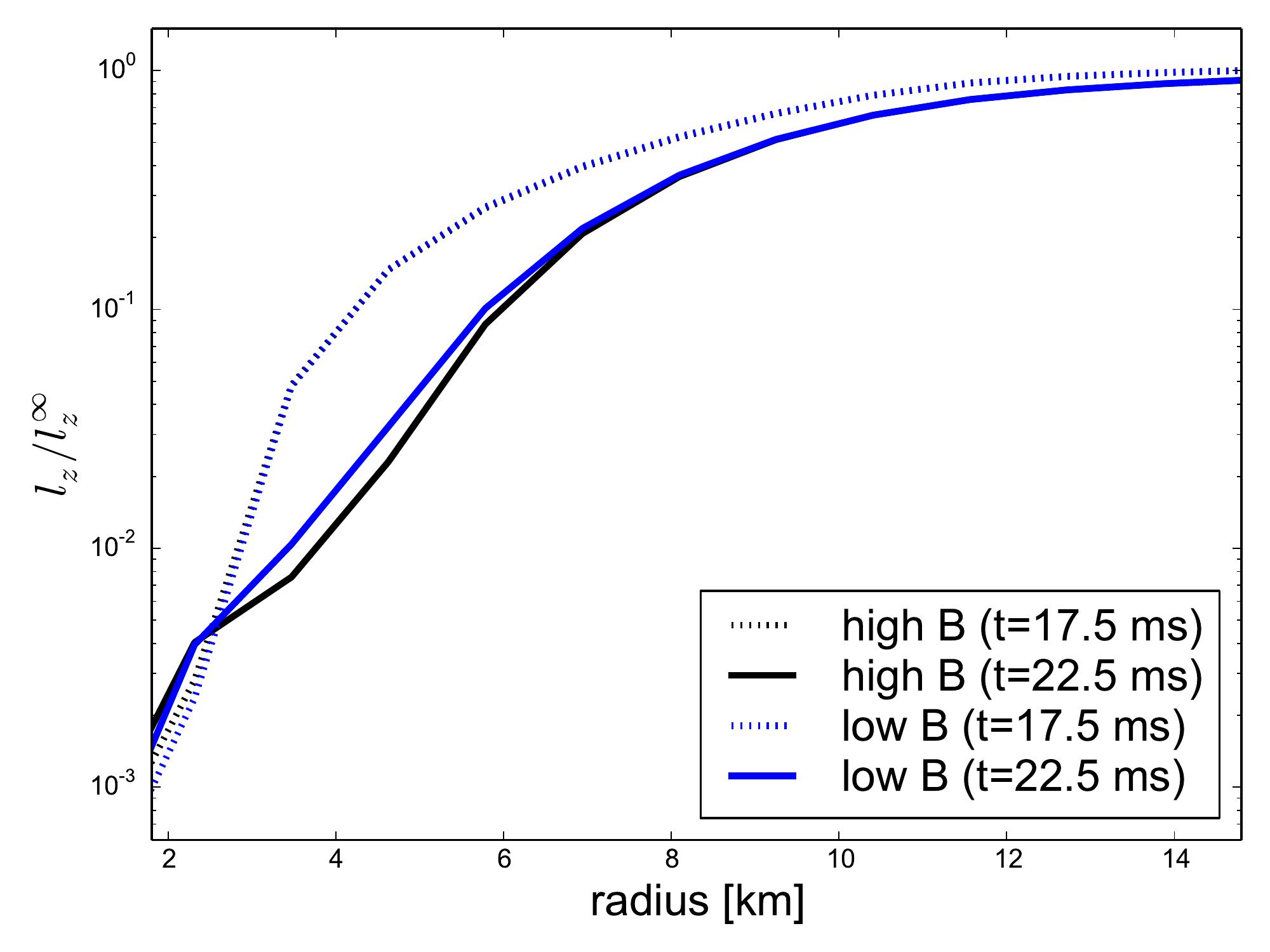}
\caption{ Comparison of the radially integrated specific angular momentum component $l_z$ for the low and high magnetized cases using the DD2 EOS at $3.2$ and $8.2$ms after the merger takes place ($t=17.5$ and $22.5$ms respectively). 
Angular momentum transport in the highly
magnetized case is stronger and, as a result, its integrated value
in the central region of the MNS is smaller than its lower magnetized counterpart case.
} 
\label{fig:magnetization_angmomcomparison}
\end{figure}

Magnetic fields can also redistribute angular momentum and transport it from
the inner to outer regions of the MNS~\cite{1999stma.book.....M}. 
Generally the processes that do this (e.g. magnetic breaking) take place over a
timescale given by $\tau \approx R_{\rm MNS}/v_A$ with $v_A$ the Alfven speed
and $R$ the radius of the star. For fields amplified to strengths
of $\approx 10^{16}$G, the time scale is roughly $\tau \approx 100$ ms and so a complete exploration of
this possibility is rather costly. Nevertheless, we can extract indications of such a redistribution
by comparing the behavior of the angular momentum for
the low and high magnetization cases. Fig.~\ref{fig:magnetization_angmomcomparison}
shows the radially integrated value of the specific, $z$-component, angular momentum, $l_z$, within
 a given radius. Apparent in the figure is a small difference between the high and low cases. In particular,
the highly magnetized case has less angular momentum at small radii than the weakly magnetized case, indicating
that angular momentum has been transported outward. As
a consequence, the contribution of rotational support against collapse weakens in the more magnetized
case which could help induce a more prompt collapse to a black hole.

We note that the atmosphere used in these two magnetized cases is larger than the unmagnetized cases by
two orders of magnitude. Because the differences that we see are largely where the outer material mixes
with the atmosphere, it becomes difficult to isolate quantitative physical effects
from atmosphere effects. Thus, the effects just discussed should be taken as qualitative indications
while we work to further improve the treatment of the atmosphere.

\section{Conclusions}
\label{sec:conclusions}
In this work we have explored the rich phenomenology of binary mergers
of equal mass neutron stars using realistic
equations of state, neutrino cooling, and electromagnetic effects. 
Our results indicate a tight connection
between different possible observables and the neutron star EoS. 
In addition to the impact of the equation of state on 
gravitational waves that are emitted during and after the merger,
both neutrino production and the properties of ejected material are strongly 
dependent on the EoS (which, in turn, would have a strong impact on electromagnetic
signals from radioactive decay).
We also developed a sub-grid model designed to capture the amplification of
the magnetic field generated by turbulent flow during the merger.
We find indications of increased angular momentum transport with the
amplified magnetic field, though
a detailed analysis is left to a future work.

The EoS imprints subtle differences on the expected GW signals prior (but close) to 
merger, and these differences become much more significant
during merger and afterward.
With softer equations of state, 
for example, the merger occurs at higher frequencies than for 
stiffer equations of state.  
The orbital frequency at merger exerts a strong influence on the resulting MNS.
In particular, an analysis of the dominant frequency components of 
the gravitational signal
after merger reveals that they can be roughly approximated by this orbital frequency 
at merger together with the characteristic oscillations of the MNS.
Our results suggest that the three waveforms presented here may 
be distinguishable by aLIGO for
optimal configurations up to distances of $\approx 120$\, Mpc (without assuming other techniques
like stacking of signals to further enhance this possibility). These results are consistent with
those found in~\cite{Read:2013zra}.

An analysis of the neutrino production indicates that
EoS yielding more compact stars (softer
EoS) produce the largest neutrino luminosity with the 
highest average neutrino energy. As discussed, this is intuitively expected as
the collision takes place deeper in the gravitational potential with a
more violent collision. We estimate that neutrinos produced by these
mergers could be observed anywhere in the Milky Way or surrounding
satellite galaxies with current neutrino detectors such as
Super-Kamiokande.  Using larger detector volumes like the proposed
Hyper-Kamiokande detector, the mergers studied here predict that we
would see $\mathcal{O}(1)$ neutrinos within the first $\approx $10-20~ms after
merger. 

An analysis of the ejecta 
indicates that only the softest
EoS yields sufficient material, and of the right quality, 
to power a kilonova event peaking in the
infrared. Indications such as this are of significant current interest in light of the recent observations reported 
in~\cite{2013Natur.500..547T,Berger:2013wna} and ~\cite{Yang:2015pha}. These observations found an infrared emission
occurring roughly a week after sGRB~130603B and about thirteen days after  GRB-060614, consistent with predictions 
of the 
decay of heavy r-process elements.  
That softer EoS merge through a more
violent collision sets the proper stage for a large amount of quite neutron rich ejecta, a requirement for producing
heavy elements through r-processes which, in turn, can produce an observable infrared emission through their radioactive decay. 

Finally, it is then interesting to connect these observations with our numerical
results to speculate about the neutron star EoS.
As has been discussed in, for example, Ref.~\cite{Piran:2014wpa}, non-vacuum
compact binaries are the prime candidates for producing heavy elements
through r-process nucleosynthesis. Such binaries are favored because of: (i)~the
mounting evidence connecting compact binary mergers with
sGRBs (e.g.~\cite{Berger:2013jza}) and (ii)~the  observational
indications~\cite{2015NatCo...6E5956W} 
that supernovae fail to produce such  heavy elements. 
If indeed these binaries are the primary producers of the heavy elements and if
sGRB~130603B and GRB~060614 are representative, then one can speculate that the
neutron star EoS is a soft one.
Simulations presented here and also in Ref.~\cite{2015arXiv150206660S} indicate that
intermediate and stiff EoS are unable to eject enough material with a
sufficiently high neutron composition, and this result, at the least, is suggestive that neutron
stars are described by a soft EoS.

However, this line of reasoning is speculative and further numerical work is needed.
In particular, the properties of ejecta resulting from the merger of unequal mass neutron
stars are yet to be analyzed, but simulations 
have shown that mergers of unequal binaries can generate accretion 
disks more massive than those produced by equal mass binaries~\cite{PhysRevLett.104.141101,2010CQGra..27k4105R}.
It thus would appear possible that the amount of 
ejecta and its properties might depend sensitively on the mass ratio. Such studies are thus an important
priority, 
although observational evidence points to small mass ratios in binary neutron star
systems~\cite{Stairs23042004}.

An alternative path could be provided by
black hole-neutron star binaries. (Ejecta estimates in such mergers have been examined recently
in~\cite{Kyutoku:2015gda,Foucart:15}.) However for sufficient ejecta with the required neutron richness: 
the black hole spin has to be quite  high, the ratio of black hole mass to neutron star mass should
be small, and the stellar EoS 
needs to be stiff. These ingredients enhance the possibility of tidal disruption. However
observed black hole masses and stellar black hole spin estimates (see e.g.~\cite{McClintock:2013vwa}),
if they hold generally, indicate black hole masses $>8 M_{\odot}$ with any possible spin.
Therefore, the likelihood of r-process nucleosynthesis occurring in black hole-neutron star binaries remains
unclear.

It is certainly quite interesting that these two possible paths---binary neutron star
or black-hole neutron star mergers---seem to prefer two very distinct possibilities
for the EoS to yield kilonova events like the ones reported. Excitingly, based on current observational evidence,
a single gravitational wave observation tied to a sGRB might be enough to single out the correct case.

\appendix
\section{The primitive solver}
\label{appendix:primsolver}

High-resolution shock-capturing schemes integrate the fluid equations
in conservation form for the conservative variables, while the fluid
equations are written in a mixture of conserved and primitive
variables. It is well known that the 
calculation of primitive variables from conserved variables
for relativistic fluids  
requires solving a transcendental set of equations. Our method for
solving these equations with a finite-temperature
EoS is a modification of the algorithm that we use
for the ideal gas EoS;
the most significant change being that the internal energy
must be calculated separately from the pressure using the table. 

The evolved conserved variables are defined as
\begin{align}
 D &\equiv \rho W \\
 S_i &\equiv \left(h W^2 + B^2\right)v_i - \left( B^j v_j\right) B_i  \\
\tau &\equiv h W^2 + B^2 - P - \frac{1}{2}\left(\left(B^i v_i\right)^2 + \frac{B^2}{W^2}\right) \\
DY_e  &\equiv \rho W Y_e.
\end{align}
The dominant energy condition places constraints on the allowed values of 
the conserved variables
\begin{equation}
D \ge 0, \quad S^2 \le (D + \tau)^2, \quad DY_e \ge 0,
\end{equation}
and depending on the EoS the second condition can be sharpened to
$ S^2 \le (2D + \tau)\tau$~\cite{Etienne:2011ea}.
These constraints may be violated during the evolution due to numerical error,
and they are enforced before solving for the primitive variables. 
A minimum allowable value of the conserved density $D_{\rm vac}$ is chosen, 
and if $D$ falls below this value we set $v^i=0$ and $D \to D_{\rm vac}$.
We choose $D_{\rm vac}$ as low as possible for the unmagnetized neutron 
star binary, which is about nine orders of magnitude smaller than the 
initial central density of the stars. For the magnetized case our solver 
fails for such tenuous
atmospheres, and it is increased by two orders of magnitude.
If the second inequality is violated, then the magnitude of $S_i$ is rescaled
to satisfy the inequality. Finally, $DY_e$ is required to satisfy the
constraint on $D$, and the computed value of $Y_e$ must be in the
equation of state table.

The primitives to be found are the density, $\rho$, pressure $P$,
electron fraction $Y_e$, internal energy density $\epsilon$ (i.e., or 
the temperature $T$, once the EoS is known) and velocity three-vector $v^i$. The magnetic field is at the same time conserved and primitive field.
We write the transcendental equations in terms of the new rescaled
variable
\begin{equation}
  x \equiv \frac{h W^2}{\rho W},
\end{equation}
where  $h$ is the {\em total} enthalpy $h \equiv \rho(1+ \epsilon) + P$,
and calculate $Y_e$ from the evolution variables  $DY_e/D$.
Following~\cite{2013PhRvD..88f4009G}, we rescale the conserved fields in order to 
get order-unity quantities, namely
\begin{equation}
  q \equiv \tau/D,\quad r \equiv S^2/D^2, \quad s \equiv B^2/D, \quad
  t \equiv B_i S^i/D^{3/2}.
\end{equation}
Then, using data from the previous time step to calculate an initial guess
for $x$, we iteratively solve these equations for $x$ within the bounds
\begin{equation}
  1 + q - s > x > 2 + 2 q - s ~~, 
\end{equation}
so that the final procedure can be written as
\begin{enumerate}
\item From the equation for $S^iS_i$, calculate an approximate Lorentz
factor $\hat{W}$ 
\[
\hat W^{-2} =
1 - \frac{x^2 r + \left( 2x + s \right) t^2}{ x^2 \left( x + s\right)^2}.
\]
\item From the definition of $D$, calculate
\[\hat \rho = \frac{D}{\hat W}.\]

\item From the definition of $\tau$ and the total enthalpy, calculate
\[
\hat \epsilon = \hat W - 1 + \frac{x}{\hat W} \left( 1 - \hat W^2 \right) + 
\hat W \left[ q - s + \frac{t^2}{2 x^2} + \frac{s}{2 \hat W^2}  \right].
\]

\item Use the EoS table to calculate the pressure $P(\hat\rho, \hat \epsilon, Y_e)$.

\item Update the guess for $x$ by solving the equation $f(x)=0$ using
the Brent method, being $f(x)$ just the definition of the unknown $x$,
\[
f(x) = x -\left(1 + \hat \epsilon 
      + \frac{P(\hat\rho, \hat \epsilon, Y_e)}{\hat \rho} \right) \hat W
\]
\end{enumerate}
The root of $f(x)=0$ from Step~5 becomes the new guess for $x$, and this
process is repeated iteratively until the solution for $x$ converges 
to a specified tolerance, which is ensured if there is a physical solution
within the bounds. One advantage of this algorithm is that $f(x)$ is
a function of a single variable, and,
in contrast to a multiple variable search for a root, robust methods
can be used to find any root that can be bracketed.

Because of numerical error, a solution to these equations may either fall
outside the physical range for the primitive variables, or a real solution
for $x$ may not exist. The solutions for  $\rho$, $T$, and $Y_e$
are, at a minimum, restricted to values in the table,
and they are reset to new values (the minimum allowed value plus ten percent) 
if necessary.  
If a real solution for the primitive variables does not exist, the primitive variables are
interpolated from neighboring points, and the conserved variables are reset
to be consistent. If a valid interpolation stencil can not 
be constructed because the solver also failed at the neighboring points,
then the update fails, and the run is terminated. 
This failure occurs very rarely and may be remedied by slightly increasing the density floor $D_{\rm vac}$.

\section{Convergence tests}
\label{appendix:convergence}

The inspiral and merger of two neutron stars is a significant test due
to the inherent asymmetry of the problem that tends to ``excite'' many,
if not all, terms in the equations. Resolving both the motion of two
compact objects as well as the large gradients at their  surfaces
require significant resources. The merger itself is a very dynamic
process with a large range of densities. Therefore, a study of the convergence of the
numerical solution is necessary to assess the validity of our results.

\begin{figure}
\centering
\includegraphics[width=8.5cm,angle=0]{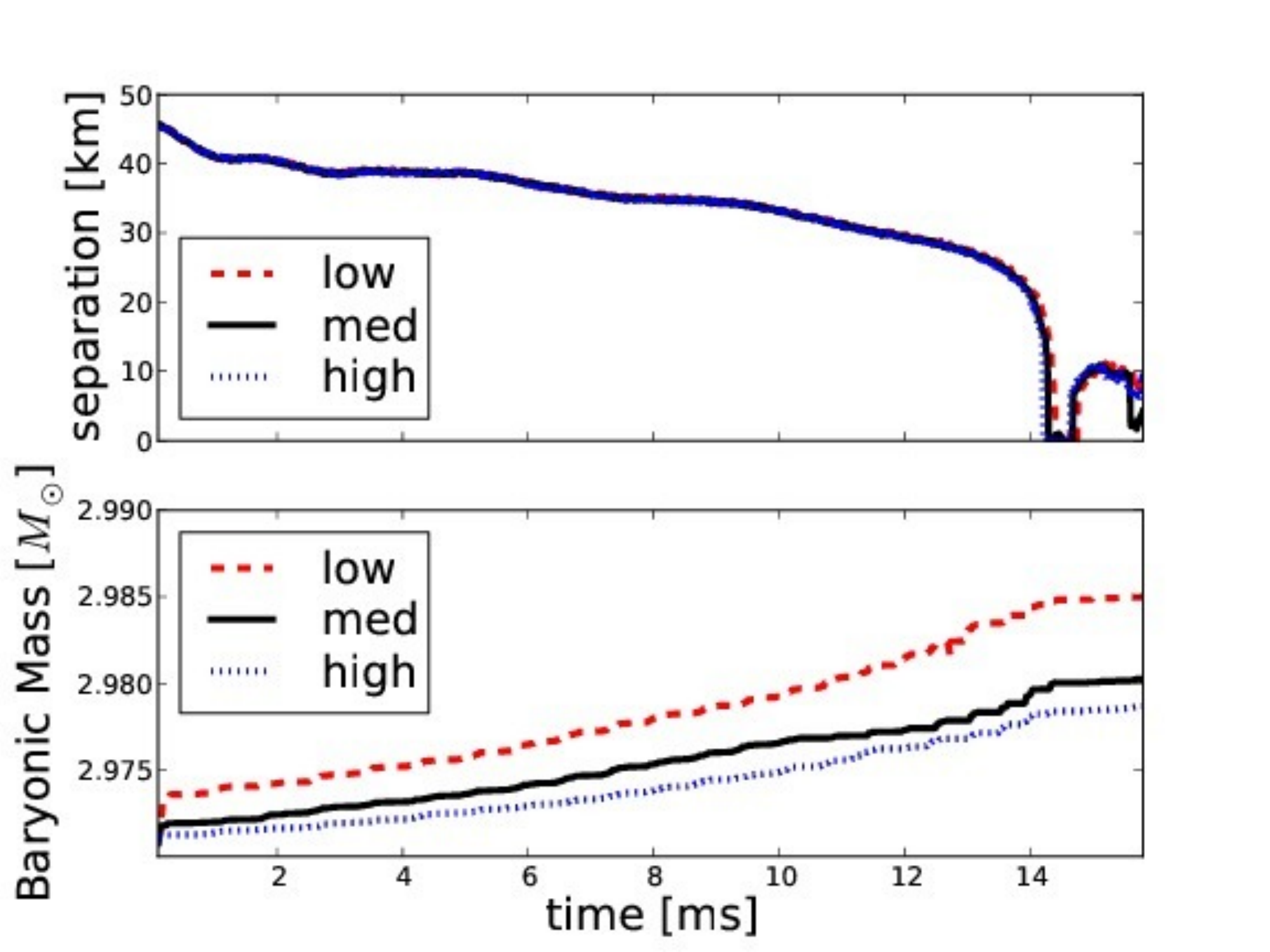}
\caption{The stellar separation and the total baryonic mass as functions
of time for the DD2 EoS with three different resolutions $\Delta x = \{ 275, 230, 192 \} \rm{m}$.
} 
\label{fig:convergece_separation}
\end{figure}

We have evolved a binary using the DD EoS 
with three different resolutions (i.e., low, medium, and high)
corresponding to $\Delta x = \{ 275,\, 230,\, 192 \}$~m. The stellar separation
and the total baryonic mass, which should
be strictly conserved during the evolution, are displayed in
Fig.~\ref{fig:convergece_separation}. Interestingly, the merger happens at earlier
times for higher resolutions, a trend opposite of that reported elsewhere. This might
indicate that the source of our numerical errors has a different sign. In any
case, the results converge at order $\approx 1.9$.

\begin{figure}[h]
\centering
\includegraphics[width=8.5cm,angle=0]{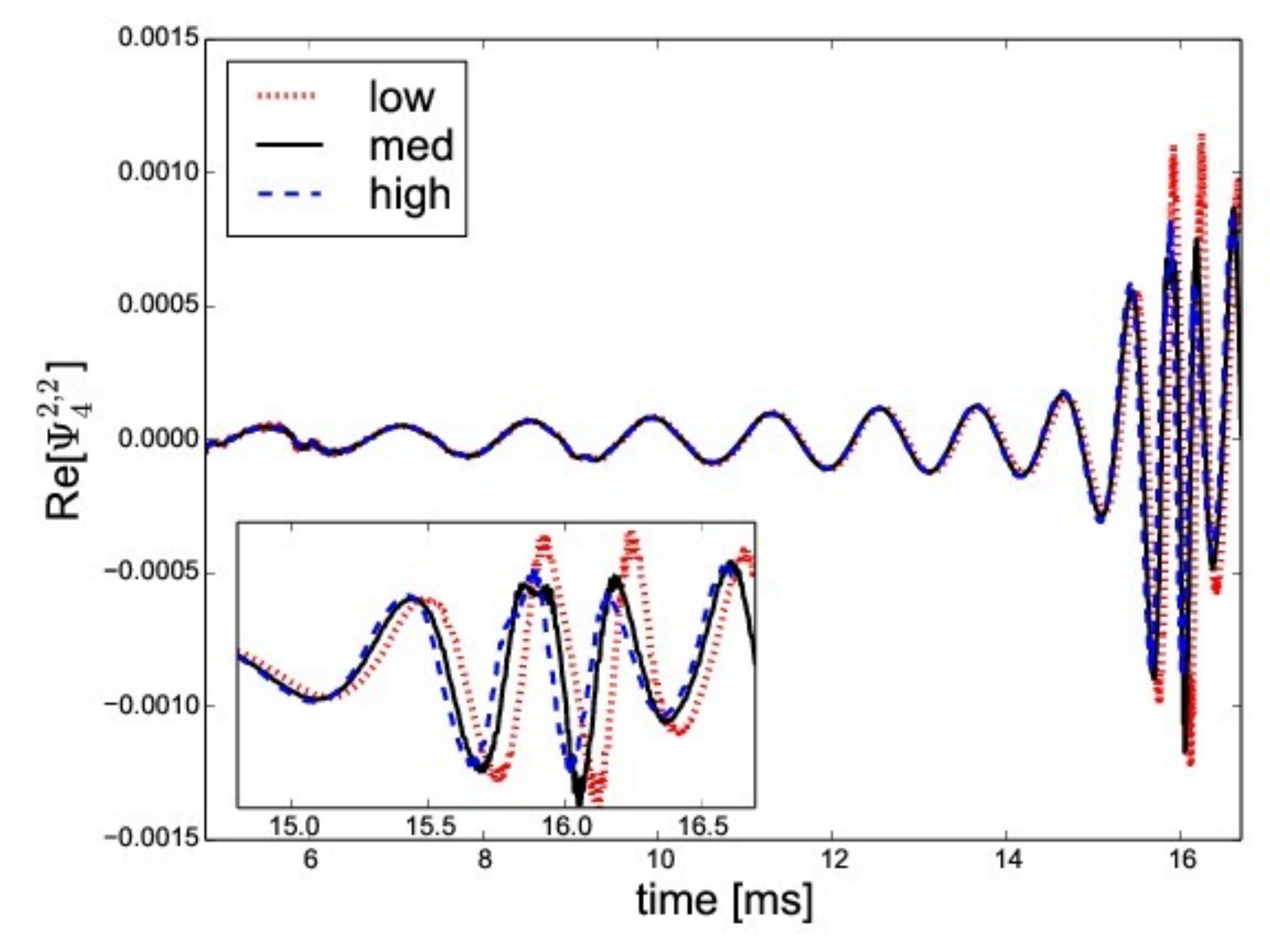}
\caption{The convergence of the principal $l=m=2$ mode of $\Psi_4$ for the same
cases as shown in Fig.~\ref{fig:convergece_separation}.
} 
\label{fig:convergence_waveform}
\end{figure}

\begin{figure}[h]
\centering
\includegraphics[width=8.5cm,angle=0]{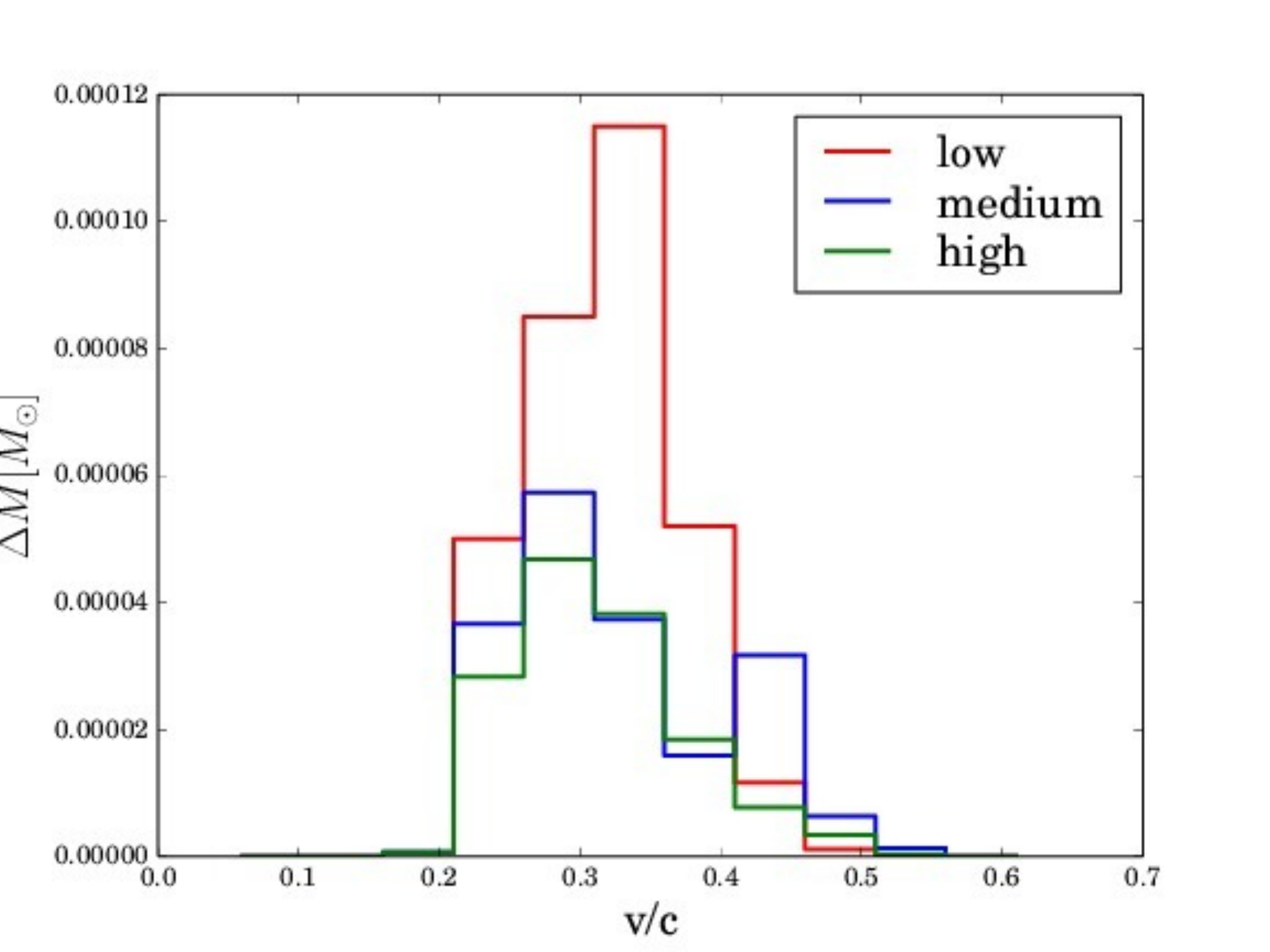}
\caption{ Histogram of the velocities, computed a few milliseconds after merger for the
same cases as shown in Fig.~\ref{fig:convergece_separation}. Notice that the medium and
high resolution results have largely converged. 
} 
\label{fig:convergence_velocities}
\end{figure}

The convergence of the principal $l=m=2$ mode of $\Psi_4$ is shown in
Fig.~\ref{fig:convergence_waveform}. Finally, the distribution of the velocity of
the fluid, computed
just few milliseconds after merger, is displayed for the different resolutions
in Fig.~\ref{fig:convergence_velocities}.

%
%
\vspace{0.5cm}

\begin{acknowledgments}
It is a pleasure to thank Jolien Creighton, Rodrigo Fernandez, Chad Hanna, Jonnas Lipuner,
Albino Perego, and Eliot Quataert for interesting discussions as well as our 
collaborators Eric Hirschmann, Patrick Motl, and Marcelo Ponce.
This work was supported by the NSF under grants PHY-1308621~(LIU),
PHY-0969811 \& PHY-1308727~(BYU), NASA's ATP program through grant NNX13AH01G,
NSERC through a Discovery Grant (to LL) and CIFAR (to LL).
CP acknowledges support from the Spanish Ministry of Education and
Science through a Ramon y Cajal grant and from the Spanish Ministry of
Economy and Competitiveness grant FPA2013-41042-P.
Additional support for this work was provided by
NASA through Hubble Fellowship grant \#51344.001-A awarded by the
Space Telescope Science Institute, which is operated by the
Association of Universities for Research in Astronomy, Inc., for NASA,
under contract NAS 5-26555.
Research at Perimeter Institute is supported through Industry Canada and 
by the Province of Ontario
through the Ministry of Research \& Innovation.  Computations were
performed at XSEDE and Scinet. 
\end{acknowledgments}

\bibliographystyle{utphys}
\bibliography{paper}

\end{document}